\documentclass[a4paper,11pt]{article}

\usepackage{jheppub} 
\usepackage[compat=1.1.0]{tikz-feynman}

\title{\boldmath Asymptotic grand unification in SO(10) with one extra dimension}

\author[a,b,c]{Gao-Xiang~Fang,}
\author[d]{Zhi-Wei~Wang}
\author[a]{and Ye-Ling Zhou}

\affiliation[a]{School of Fundamental Physics and Mathematical Sciences, Hangzhou Institute for Advanced
Study, UCAS, Hangzhou 310024, China}
\affiliation[b]{Institute of Theoretical Physics, Chinese Academy of Sciences, Beijing 100190, China}
\affiliation[c]{University of Chinese Academy of Sciences, Beijing 100049, China}
\affiliation[d]{School of Physics, The University of Electronic Science and Technology of China, 
No.~2006, Xiyuan Avenue, West Hi-Tech Zone, Chengdu, China}

\emailAdd{fanggaoxiang21@mails.ucas.ac.cn}
\emailAdd{zhiwei.wang@uestc.edu.cn}
\emailAdd{zhouyeling@ucas.ac.cn}

\abstract{
Asymptotic grand unification provides an alternative approach to gradually unify gauge couplings in the UV limit, where they reach a non-trivial UV fixed point. 
Using an economical and realistic particle content setup, we demonstrate that asymptotic grand unification can be achieved in a 5D SO(10) model with one extra dimension. The top, bottom and tau masses are split, and the smallness of the neutrino mass is explained via inverse seesaw.
One intermediate scale, the Pati-Salam symmetry breaking scale, is included below the compactification scale. Due to the absence of large-dimensional Higgs representations, gauge couplings exhibit asymptotic safety and are thus asymptotically unified, regardless of their initial values. In contrast, Yukawa couplings can achieve asymptotic freedom if the negative gauge contributions dominate over the positive Yukawa terms, requiring exact unification at the compactification scale. The widely-used 126-dimensional Higgs is not recommended in this 5D asymptotic SO(10) GUT, as it tends to drive the gauge beta function positive, compromising asymptotic safety.
}


\begin{document} 
\maketitle
\flushbottom

\section{Introduction}

Unification is an attractive concept in particle physics since it successfully unifies the electromagnetic interaction and weak interaction in the semi-simple gauge group $SU(2)_L \times U(1)_Y$. Pati and Salam (PS) attempted to unify quarks and leptons in $SU(4)_c \times SU(2)_L \times SU(2)_R$ by regarding leptons as the fourth colour of quarks \cite{Pati:1973uk,Pati:1974yy}. The first grand unified theory (GUT) is proposed by George and Glashow (GG), where all known gauge interactions are unified in a simple Lee group $SU(5)$ \cite{Georgi:1974sy}. Later, Fritzsch and Minkowski proposed $SO(10)$ GUT \cite{Fritzsch:1974nn}, which contains PS and GG, as well as flipped SU(5) \cite{Barr:1981qv, Derendinger:1983aj} as its sub-symmetries. 

Asymptotic unification, distinct from the conventional notion of unification, suggests that couplings unify at a non-trivial ultraviolet (UV) fixed point at an infinitely high energy scale. Theories with an interacting UV fixed point is normally referred to as asymptotic safety which was initially proposed by S.~Weinberg in \cite{Weinberg:1979} to address the famous UV Landau pole problem. For asymptotic safety involving gravity please refer to a review e.g.~\cite{Bonanno:2020bil} and in this work, however, we focus on safety without gravity. The first asymptotically safe 4-dimensional non-supersymmetric gauge-Yukawa theory without gravity was established in \cite{Litim:2014uca} where Veneziano limit (i.e.~both large number of colour and large number of flavours) is required. Later, an alternative framework aiming to realize asymptotic safety of the Standard Model (SM) with large number of flavours $N_f$ was proposed in \cite{Mann:2017wzh} and further lead to the proposal of UV conformal window in \cite{Antipin:2017ebo}. The details of this large $N_f$ framework are nicely summarized in \cite{Antipin:2018zdg,Pelaggi:2017abg,Kowalska:2017pkt}. The requirement of the existence of an UV fixed point helps to resolve the crisis that GUTs especially based on supersymmetry may feature a Landau pole right above the unification scale due to a large number of matter fields. This motivates the initial idea of asymptotically safe GUT appeared in \cite{Bajc:2016efj} which focuses on a formal consistency check of an UV fixed point. Explicit examples of asymptotically safe Pati-Salam models and Trinification models within the large $N_f$ framework can be found respectively in \cite{Molinaro:2018kjz,Sannino:2019sch} and \cite{Wang:2018yer}. The above asymptotically safe GUT works, in our opinions, are more focusing on the existence of an UV fixed point in a GUT (or semi-simple GUTs) rather than how the interaction couplings could unify asymptotically. Thus, they can be viewed as a pre-version of the real asymptotic unification.

The first concrete realization of asymptotic unification (or asymptotic GUT/aGUT) based on the minimal $SU(5)$ model in 5D is achieved in \cite{Cacciapaglia:2020qky}, followed by a supersymmetric exceptional $E_6$ case in \cite{Cacciapaglia:2023ghp}, and a systematic classification of $SU(N)$ theories in \cite{Cacciapaglia:2023kyz}, where by definition, all the couplings are unified asymptotically at a non-trivial UV fixed point. 
Beyond the intriguing feature of unification in the UV limit, the existence of such a fixed point in extra dimensions also implies that the theories are renormalizable, in contrast to the common view that quantum field theories in extra dimensions are merely effective theories \cite{Gies:2003ic,Morris:2004mg,Pastor-Gutierrez:2022rac,Pastor-Gutierrez:2024huj}.
Contrary to the standard unification, asymptotic GUTs are driven by the contributions of complete multiplets of the GUT symmetry in the bulk. Unlike the above 4D asymptotic safety scenarios, where UV fixed points arise due to intricate gauge-Yukawa interplay or pole structure in the large $N_f$ resummation function, the UV fixed point in 5D theories emerges from the contributions of Kaluza-Klein (KK) states to the running couplings when above the KK scale $M_{\rm KK}$, which also called the compactification scale, i.e.~when the compactified extra dimension becomes relevant \cite{Dienes:1998vh,Dienes:2002bg}. In this framework, the extra dimension is compactified through orbifolding, which consistently breaks the bulk gauge symmetry. Orbifolding refers to the identification of points in the extra dimension under a discrete symmetry, typically resulting in specific symmetry-breaking loci (commonly called orbifold fixed point\footnote{These orbifold fixed point should not be confused with the ultraviolet fixed point mentioned in the text.}) where the symmetry is reduced. A recent study regarding the orbifolding stability can be found in \cite{Cacciapaglia:2024duu,Cacciapaglia:2025bxs}.
From a 5D perspective, the gauge coupling flows to an interacting UV fixed point, whereas, in the corresponding 4D picture, it runs to a Gaussian fixed point with zero value, exhibiting asymptotic freedom. This asymptotic freedom characteristic in 4D imposes a limit on the maximum amount of matter contents that can be included; exceeding this limit would result in a Landau pole in the gauge $\beta$ functions. This constraint serves as a crucial criterion/guidline for selecting matter contents during model construction. In contrast, the 4D safety scenarios mentioned above impose no such upper bound on the matter content. To ensure the total asymptotic safety of the system—where all couplings are required to flow to a fixed point—the Yukawa couplings must also run into either an interacting fixed point (safe) or a non-interacting Gaussian fixed point (free), alongside satisfying the constraints from the gauge $\beta$ functions. In \cite{Khojali:2022gcq}, the authors applied the framework to $SO(10)$ with an overly simplified Higgs sector, leading the Yukawa coupling to flow to a Landau pole rather than a fixed point. However, to fully establish the viability of the asymptotic unification scenario for the $SO(10)$ case, a more comprehensive analysis, particularly involving a more complete Higgs sector, is necessary.

For a realistic GUT model, it is essential to consider a complete Higgs sector.  While some of the Higgs fields are necessary to achieve the spontaneous breaking of GUT symmetry as well as the intermediate gauge symmetries, additional Higgs fields are often motivated by practical phenomenological considerations. The first motiviation is the need to split the masses of the $b$ quark and the $\tau$ lepton, which typically requires more than one Higgs multiplet \cite{Glashow:1979nm,Magg:1980ut,Mohapatra:1980yp,Lazarides:1980nt,Senjanovic:1981ff} or the inclusion of higher dimensional operators \cite{Ellis:1979fg,Weinberg:1979sa,Shafi:1983gz}. The second, and what is more important in our opinion, is to recover all known experimental data on the masses and mixing of quarks and leptons, in particular, accounting for neutrino oscillation data \cite{Dutta:2004hp,Dutta:2005ni,Dutta:2004zh,Grimus:2006bb,Grimus:2006rk}. Efforts to fit the increasingly precise data in the $SO(10)$ GUT framework have been explored in \cite{Joshipura:2011nn, Altarelli:2013aqa, Dueck:2013gca, Babu:2016bmy, Fu:2022lrn, Fu:2023mdu}, all of which include more than one copy of Higgs multiplets. From a phenomenological perspective, it is therefore crucial to verify whether an asymptotic GUT model can realistically accommodate these requirements.

In this work, we discuss the asymptotic GUT in $SO(10)$ with one extra dimension. While recovering all experimental data is beyond the scope of this study, we introduce additional Higgs multiplets at the GUT scale to ensure a realistic particle spectrum. These Higgs multiplets play a crucial role in ensuring the asymptotic safety of Yukawa couplings in the UV limit. The rest of the paper is organised in the following. We give the general setup of 5D $SO(10)$ GUT in section~\ref{sec:2}. Section~\ref{sec:3} discusses the asymptotic behavior and unification of gauge couplings. Section~\ref{sec:4} provides a detailed analysis of the renormalization group (RG) running and the asymptotic behaviour of Yukawa couplings. We conclude in section~\ref{sec:5}. Renormalization group equation (RGE) for Yukawa couplings in both 4D and 5D are derived in Appendices~\ref{app:2} and \ref{app:3} for reference. We further show RGE matching between PS and SO(10) in Appendix~\ref{app:4}. 

\section{The framework}\label{sec:2}

\begin{table}[h]
	\begin{center}
		\begin{tabular}{ c c c c}
			\hline \hline
			Energy scale & Symmetry & Fermion & Higgs  \\
			\hline
			$\mu> M_{\rm KK}$ & $SO(10)$ & 
			$\begin{array}{c}\Psi_{\bf 16} \sim {\bf 16}\\ 
			\Psi_{\overline{\bf 16}}\sim \overline{\bf 16} \\
			\nu_{\rm S} \sim {\bf1} \end{array}$ & 
			$\begin{array}{c}H_{\bf10} \sim {\bf10}, \text{complex}\\ 
			H_{\bf120} \sim {\bf120}, \text{real} \\
			H_{\bf16} \sim{\bf16} \\
\end{array}$ \\\hline
             $M_{\rm PS}<\mu< M_{\rm KK}$ & $G_{\rm 422}$ & 
             $\begin{array}{c}\psi_L \sim (4,2,1) \\ 
             \psi_R \sim (4,1,2) \\ 
             \nu_{\rm S} \sim (1,1,1)
             \end{array}$ & 
             $\begin{array}{c} h_1,h_1' \sim (1,2,2), h_{15}\sim(15,2,2) \\ 
              h_{\bar4}\sim (\bar4,1,2) \end{array}$ \\\hline
             $M_{Z} \ll \mu< M_{\rm PS}$ & $G_{\rm SM}$ &
             $\begin{array}{c}q_L,d_R,u_R \\ l_L,\nu_R, e_R, \nu_{\rm S} \end{array} $ & 
             $h_{\rm SM}$ 
             \\\hline \hline
		\end{tabular}
	\end{center}
	\caption{Gauge symmetries and particle contents remnant of the model at different energy scales. $\mu \gg M_{Z}$ means $\mu$ is at a sufficiently high scale such that heavy neutrinos has not decoupled. $G_{422} \equiv SU(4)_c \times SU(2)_L \times SU(2)_R,G_{\rm SM}\equiv SU(3)_c\times SU(2)_L\times U(1)_Y$, where $q_L=(u_L,d_L)$, $l_L=(\nu_L,e_L)$, $q_R = (u_R, d_R)$, $l_R = (\nu_R, e_R)$. $H_{\bf 10}$ and $H_{\bf 120}$ can be either real or complex. For an economical realization of fermion masses, we assume $H_{\bf 10}$ to be complex and  $H_{\bf 120}$ to be real throughout this work.
	\label{tab:particle_contents}}
\end{table}

We construct a realistic $SO(10)$ grand unified theory in 5-dimensional spacetime, where the extra dimension is compactified on an $S_1/(Z_2 \times Z_2')$ orbifold \cite{Meloni:2016rnt, Ohlsson:2018qpt} with length scale $L$. Orbifold compactification of a single extra dimension allows large grand unified symmetries to be broken down to their subgroups without reducing the rank of the group \cite{Hebecker:2001jb}. Based on this framework, we consider the following breaking chain:
\begin{eqnarray} \label{eq:breaking_chain}
{\rm 5D}\ SO(10) \underset{\text{BC}}{\xrightarrow{M_{\rm KK}}} \text{4D Pati-Salam}
\underset{{\bf 16}}{\xrightarrow{M_{\rm PS}}} \text{SM} \,,
\end{eqnarray}
where BC, $M_{\rm PS}$ and $\bf 16$ denote respectively the boundary conditions, the Pati-Salam symmetry breaking scale as well as the Higgs in the spinor representation $H_{\bf 16}$ in Table.~\ref{tab:particle_contents}. In our extra-dimensional $SO(10)$ model, the chiral SM fermions are identified as the zero modes of bulk fields. Due to the boundary conditions that break $SO(10)$, it is not possible to embed the entire set of SM fermions with zero modes within a single multiplet. We therefore introduce $\Psi_{\bf 16}$ and $\Psi_{\overline{\bf 16}}$ to accommodate the SM fermions. The matter fields decompose as $\Psi_{\bf 16}  = \psi_L + \Psi_R^c$ and $\Psi_{\overline{\bf 16}} = \Psi_L^c + \psi_R$ after the gauge symmetry is broken to $G_{422}$ at the KK scale. Here
\begin{align}
\psi_L&=(q_L, l_L) \sim(4, 2, 1)\,,\quad\;\;
\Psi_R^c =(Q_R^c, L_R^c) \sim (\overline{4}, 1, 2) \,, \nonumber\\
\Psi_L^c& =(Q_L^c, L_L^c)\sim (\overline{4}, 2, 1) \,, \quad 
\psi_R = (q_R, l_R) \sim(4, 1, 2)\,,\quad
\end{align}
where the capitalized letters such as $Q,\,L$ represent fields without zero modes. This arrangement forbids proton decay as all the field components without a zero mode have $B$ and $L$ values which are half of those of the SM particles \cite{Cacciapaglia:2020qky}. 

To generate fermion mass splittings between quarks and leptons, we introduce two Higgs multiplets, ${\bf 10}$ and ${\bf 120}$. In the minimal setup, we take $H_{\bf 10}$ to be complex and $H_{\bf 120}$ to be real to split fermion masses, even though both ${\bf 10}$ and ${\bf 120}$ are real representations of $SO(10)$. This point will be clarified later in the section. A Higgs field in the chiral represenation, $H_{\bf16}$, which contains a trivial SM singlet $h_{\rm S}$, is introduced to break the Pati-Salam symmetry to the Standard Model. A sterile neutrino $\nu_{\rm S}$, also an $SO(10)$ singlet, is introduced as the source of Majorana masses for light neutrinos. In this work, we have the following matter fields and Higgs fields,
\begin{align}
\Psi_{\bf16}, \Psi_{\overline{\bf16}}, \nu_{\bf S}, H_{\bf10}, H_{\bf120}, H_{\bf16} \,.
\end{align}
A distinctive feature of our model, compared to conventional $SO(10)$ GUTs, is the absence of the usual $\overline{\bf126}$-plet Higgs. As we will show later, this is crucial for avoiding a Landau pole in the Yukawa coupling.

In order to implement the desired boundary conditions, we assign two parities $P_0$ and $P_1$, associated with the two fixed points $y=0$ and $y=L$ in the 10-dimensional fundamental representation of $SO(10)$ as
\begin{align}
P_0=&\left(+\,+\,+\,+\,+\,+\;-\,-\,-\,-\right)\,, \nonumber\\
P_1=&\left(+\,+\,+\,+\,+\,+\;+\,+\,+\,+\right)\,.
\end{align}
More precisely, $P_0$ and $P_1$ acts on the fundamental representation $\phi$ at the UV and IR branes as 
\begin{align}
& \hat{P}_0 \phi(x,y) \hat{P}_0^{-1} = \phi(x,-y) = P_0 \phi(x,y) \,, \nonumber\\
& \hat{P}_1 \phi(x,y) \hat{P}_1^{-1} = \phi(x,2 L-y) = P_1 \phi(x,y)\,.
\end{align}
Imposing parity invariance, the gauge bosons, which belong to the adjoint representation of $SO(10)$, are assigned boundary conditions as follows: the components $A_\mu$ have BCs $(+,+)$ and $A_y$ have BCs $(-,-)$ in the upper-left $6\times 6$ block and the lower-right $4\times 4$ blocks. The off-diagonal blocks are assigned BCs $(-,+)$ for $A_\mu$ and BCs $(+,-)$ for $A_y$. These boundary conditions result in the breaking of $SO(10)$ to $SO(6)
\times SO(4)$, which is locally Isomorphic to $SU(4)_c \times SU(2)_L \times SU(2)_R$.  
BSs for fermions and Higgs are arranged as follows: fields denoted with lowercase letters under the PS or SM gauge symmetries correspond to fields with zero modes, while those labeled with uppercase letters correspond to those with KK modes only. A complete list of these fields is provided in Table~\ref{tab:particle_contents_2} in Appendix~\ref{app:1}, where their BCs can be arranged properly following the lowercase or uppercase notatioin.


The Yukawa coupling terms in 5D $SO(10)$, following the usual 4D left-right convention (i.e., $m\,\overline{\psi_L}\, \psi_R$), take the form
\begin{align}
		-\mathcal{L}_Y=&y_{\bf 10} \overline{\Psi_{\bf 16}} H_{\bf 10} \Psi_{\overline{\bf 16}} + {\rm i} y_{\bf 120} \overline{\Psi_{\bf 16}} H_{\bf 120} \Psi_{\overline{\bf 16}} + y_{\bf 16} \overline{\nu_{\rm S}} H_{\bf 16} \Psi_{\overline{\bf 16}} + \frac{L}{2} \mu_{\rm M} \, \overline{\nu_{\rm S}} \nu_{\rm S}^c\, \delta(y-L) + {\rm h.c.}\,.
		\label{eq:Lagrangian_SO10}
\end{align}
There might be additional terms $\overline{\Psi_{\bf 16}} H_{\bf 10}^* \Psi_{\overline{\bf 16}}$ for the complex field $H_{\bf 10}^*$. Including such a term is not essential to achieve the UV safety of Yukawa coupling but just increases the complexity of Yukawa RG running discussed later in section~\ref{sec:4}, and thus we will forbid it by assuming a global $U(1)$ symmetry. 
The Majorana mass term is localized on the IR brane ($y = L$), following the approach of Ref.~\cite{Alezraa:2025lct}, where $\mu_{\rm M}$ denotes the Majorana mass parameter.
For simplicity, we focus only on the heaviest family in the main text.\footnote{In conventional 4D SO(10) models, all SM chiral fermions are embedded in the same ${\bf16}$ spinor representation. Following the decomposition of product, ${\bf16} \times {\bf16} = {\bf10}_{\rm s} + {\bf126}_{\rm s} + {\bf120}_{\rm a}$, three Higgs multiplets $H_{\bf10}$, $H_{\overline{\bf126}}$, $H_{\bf120}$ are introduced, and the Yukawa couplings are arranged as 
\begin{align}
{\bf16}_F^\alpha (Y_{\bf10}^{\alpha\beta} H_{\bf10} + Y_{\overline{\bf126}}^{\alpha\beta} H_{\overline{\bf126}} + {\rm i} Y_{\bf120}^{\alpha\beta} H_{\bf120}) {\bf16}_F^\beta + {\rm h.c.},
\end{align}
where $\alpha,\beta$ are flavour indices. The Yukawa matrices in the flavour space satisfy $Y_{\bf10}^{\alpha\beta} = Y_{\bf10}^{\beta\alpha}$ and $Y_{\overline{\bf126}}^{\alpha\beta} = Y_{\overline{\bf126}}^{\beta\alpha}$ and $Y_{\bf120}^{\alpha\beta} = - Y_{\bf120}^{\beta\alpha}$. These symmetric or antisymmetric property are restricted by the symmetric (${\bf10}$ and ${\bf126}$) and antisymmetric (${\bf120}$) construction in the gauge space. In particular, if we assume just one family, $Y_{\bf120}$ = 0. However, these symmetry constraints do not apply in this work since the SM left-handed fermions and right-handed fermions reside in different spinor representations $\Psi_{\bf 16}$ and $\Psi_{\overline{\bf 16}}$. All Yukawa matrices in the flavour space are therefore unconstrained in the flavour space. In this work, since we consider only the heaviest family, both $Y_{\bf10}$ and $Y_{\bf120}$ reduce to numbers, denoted by $y_{\bf10}$ and $y_{\bf120}$.}  Then, all Yukawa couplings $y_{\bf 10}$, $y_{\bf 120}$, and $y_{\bf 16}$ are treated as numbers and we further assume them to be real to reduce the number of free parameters. 
The $SO(10)$ algebra and convention of Yukawa couplings are introduced in Appendix~\ref{app:3}. 
Below the KK scale, the zero-mode fields lead to the following Yukawa interactions
\begin{align}
		-\mathcal{L}_Y^{\rm 4D}\supset  y_1 \overline{\psi_L} h_1 \psi_R +  \overline{\psi_L} (y_1' h_1'+ y_{15} h_{15}) \psi_R + y_4 \overline{\nu_{\rm S}} h_{\bar4} \psi_R + \frac{1}{2}\mu_{\rm M} \overline{\nu_{\rm S}} \nu_{\rm S}^c + {\rm h.c.}\,,
		\label{eq:Lagrangian_PS}
\end{align}
where 
\begin{align}
   y_1 = \sqrt{2} \, y_{\bf 10} \,,\quad
   y_1' = \sqrt{2} y_{\bf 120} \,,\quad
   y_{15} = 2 \sqrt{2} y_{\bf 120} \,,\quad
   y_4 = y_{\bf 16}\,. \label{eq:matching}
\end{align}

The Higgs fields $h_1$, $h_1'$ and $h_{15}$ further decompose into electroweak doublets once the PS symmetry is broken to the SM gauge group. These doublets mix, and the SM Higgs emerges as the lightest, massless eigenstate. Once the SM 
Higgs gains a VEV, these doublets in $H_{\bf 10},H_{\bf 120}$ share fractions of the VEV, which are denoted as
\begin{align} 
	\{\langle h_u \rangle, \langle h_d \rangle, \langle h'_u \rangle, \langle h'_d \rangle, \langle h''_u \rangle, \langle h''_d \rangle \} = \{c^u_{\bf10},c^d_{\bf10}, c^{d'}_{\bf120},c_{\bf120}^{d'}, c_{\bf120}^d,c_{\bf120}^d \} \times v_{\rm EW} \,,
	\label{eq:vev}
\end{align}
with $v_{\rm EW}=174\ {\rm GeV}$ and $\sum |c|^2 = 1$. The resulting Dirac mass matrices for fermions take the form $m_f = y_f v_{\rm EW}$ (for $f=t,b,\tau$) and $m_{\rm D} = y_\nu v_{\rm EW}$ for the neutrino Dirac mass, with
\begin{align}
		&y_t= \sqrt{2}\, y_{\bf 10}c_{\bf10}^u+\sqrt{2}\, y_{\bf 120}(c_{\bf120}^{d'}+ \frac{1}{\sqrt{3}} c_{\bf120}^d)\,, \nonumber\\
		&y_b=\sqrt{2}\, y_{\bf 10}c_{\bf10}^d+\sqrt{2}\, y_{\bf 120}(c_{\bf120}^{d'}+ \frac{1}{\sqrt{3}} c_{\bf120}^d)\,, \nonumber\\
		&y_\tau=\sqrt{2}\, y_{\bf 10}c_{\bf10}^d+\sqrt{2}\, y_{\bf 120}(c_{\bf120}^{d'}- \sqrt{3} c_{\bf120}^d)\,, \nonumber\\
		&y_\nu=\sqrt{2}\, y_{\bf 10}c_{\bf10}^u+\sqrt{2}\, y_{\bf 120}(c_{\bf120}^{d'}- \sqrt{3} c_{\bf120}^d)\,.
	\label{eq:SO10mass}
\end{align}
Following the decomposition in Eq.~\eqref{eq:Lagrangian_PS} and the matching conditions in Eq.~\eqref{eq:matching}, the Yukawa couplings can be re-written in terms of the Pati-Salam gauge theory as:\begin{align}
		&y_t=y_{1}c_{\bf10}^u+y'_{1}c_{\bf120}^{d'}+\frac{1}{2\sqrt{3}}y_{15}c_{\bf120}^d\,, \nonumber\\
		&y_b=y_{1}c_{\bf10}^d+y'_{1}c_{\bf120}^{d'}+\frac{1}{2\sqrt{3}}y_{15}c_{\bf120}^d\,, \nonumber\\
		&y_\tau=y_{1}c_{\bf10}^d+y'_{1}c_{\bf120}^{d'}-\frac{\sqrt{3}}{2}y_{15}c_{\bf120}^d\,, \nonumber\\
		&y_\nu=y_{1}c_{\bf10}^u+y'_{1}c_{\bf120}^{d'}-\frac{\sqrt{3}}{2}y_{15}c_{\bf120}^d\,.
	\label{eq:Yukawa_PS}
\end{align}
In this section, we will only consider the tree-level matching of Yukawa couplings. The effects of Yukawa RG running will be addressed in section~\ref{sec:4}. 

We further comment on the assumption that $H_{\bf 10}$ is complex and $H_{\bf 120}$ is real. This choice provides the minimal set of Higgs degrees of freedom necessary to split the Yukawa couplings of the top and bottom quarks and tau lepton. 
If both $H_{\bf 10}$ and $H_{\bf 120}$ are taken to be complex, Eq.~\eqref{eq:vev} generalizes to 
\begin{align} 
	\{\langle h_u \rangle, \langle h_d \rangle, \langle h'_u \rangle, \langle h'_d \rangle, \langle h''_u \rangle, \langle h''_d \rangle \} = \{c^u_{\bf10},c^d_{\bf10}, c^{u'}_{\bf120},c_{\bf120}^{d'}, c_{\bf120}^u,c_{\bf120}^d \} \times v_{\rm EW} \,,
	\label{eq:vev_p}
\end{align} 
and Yukawa couplings are modified accordingly. The two additional parameters $c^{u'}_{\bf120}$ and $c^{u}_{\bf120}$ further relax the correlations between quark and lepton Yukawas. However, this case leads to an increased UV fixed point value of gauge couplings, which will be discussed in the next section; see Table~\ref{tab:fixed_points}. 
Conversely, if both $H_{\bf 10}$ and $H_{\bf 120}$ are taken to be real, we arrive at an over-simplified version with coefficients $c_{\bf10}^u = c_{\bf10}^d$ in Eq.~\eqref{eq:vev}, leading to $y_t = y_b$, which is incompatible with experimental data. Therefore, we adopt the assumption that $H_{\bf 10}$ is complex and $H_{\bf 120}$ is real throughout this work.

Moreover, the Yukawa Lagrangian in Eq.~\eqref{eq:Lagrangian_SO10} contains the terms $y_{16}\, \overline{\nu_{\rm S}} h_{\rm S} \nu_R$. Once $h_{\rm S}$ gains the VEV, which are denoted as
	$\langle h_{\rm S}\rangle = v_{\rm S}$,
this term contributes to the mass term $m_{\rm S}  \overline{\nu_{\rm S}} \nu_R$
with $m_{\rm S}=y_{16}v_{\rm S}$. Note that the scale $v_{\rm S} \sim M_{\rm PS}$ also drives the symmetry breaking of the 4D Pati-Salam symmetry down to the SM. The mass parameters $m_{\rm D},m_{\rm S},\mu_{\rm M}$ then enter the neutrino mass matrix, expressed in the symmetric basis $(\nu_L^c,\nu_R,\nu_{\rm S}^c)$ as:
\begin{eqnarray}
	\left(
	\begin{array}{ccc}
		0 & m_{\rm D} & 0 \\
		m_{\rm D} & 0 & m_{\rm S} \\
		0 & m_{\rm S} & \mu_{\rm M}
	\end{array}
	\right).
	\label{eq:neutrino_matrix}
\end{eqnarray}
After the diagoinalisation of the matrix,  the light neutrinos masses arise via the inverse seesaw mechanism \cite{Wyler:1982dd, Mohapatra:1986bd}, leading to the effective mass as 
\begin{eqnarray}
	m_\nu=  \mu_{\rm M} \frac{m_{\rm D}^2}{m_{\rm S}^2} \,.
	\label{eq:effective_neutrino_matrix}
\end{eqnarray}
A general parametrisaton of the symmetric $9\times 9$ Majorana mass matrix referring to the 3-family extension of Eq.~\eqref{eq:neutrino_matrix} is given in  \cite{Han:2021qum}. 




\section{Gauge running and asymptotic unification} \label{sec:3}

Given the symmetry breaking chain outlined in Eq.~\eqref{eq:breaking_chain}, we calculate the RG running of the fundamental couplings from the electroweak (EW) scale up to the Planck scale. This section focuses on the RG running of gauge couplings, while the next section will address the running of Yukawa couplings. 

For a simple Lie group  $H_i$, the one-loop RGE for its gauge coupling $g_i$ is given by $ (16\pi^2) {\rm d} g_i / {\rm d} t = b_i g_i^3$, where the RG time is defined as $t = \ln \mu / \mu_0$ with $\mu_0 = M_Z$ and $\mu> M_Z$.  It is convenient to re-write it in terms of $\alpha_i=g_i^2/4\pi$, yielding
\begin{eqnarray} \label{eq:RGE_gauge1}
 2\pi \frac{{\rm d} \alpha_i}{{\rm d} t} = b_i \alpha_i^2 \,,
\end{eqnarray}
where the RG $\beta$-coefficient $b_i$ is:
\begin{eqnarray}
	b_i = - \frac{11}{3} C_2(H_i) + \frac{2}{3} \sum_F T(F_i) + \frac{1}{6} \sum_S T(S_i),
\end{eqnarray}
with $F$ and $S$ denoting all chiral fermion and real scalar multiplets respectively while $F_i$ and $S_i$ are their respective representations under the group $H_i$. The term $C_2(R_i)$ (with $R_i=F_i, S_i$) represents the quadratic Casimir of the representation $R_i$, while $C_2(H_i)$ corresponds to that of the adjoint representation of the group $H_i$. 

For the RG evolution from the EW scale $M_Z$ to the PS scale $M_{\rm PS}$, the three gauge couplings evolve as follows: $g_3$ for $SU(3)_c$, $g_{2L}$ for $SU(2)_L$ and $g_1 = \sqrt{5/3} g_Y$ for $U(1)_Y$.  The corresponding one-loop $\beta$-coefficients are: 
\begin{eqnarray}
(b_3^{\rm SM},b_{2L}^{\rm SM},b_1^{\rm SM})= \Big(-7,\; -\frac{19}{6},\; \frac{41}{10} \Big) \,.
\end{eqnarray}
In this sector, only the SM Higgs has been included. Additional Higgs multiplets, decomposed after the PS breaking, may in principle contribute to the running; however, we have neglected the potential threshold effects, which might become important if their masses deviate substantially from the PS scale. 

For the RG running from $M_{\rm PS}$ to the compactification scale $M_{\rm KK}$, the gauge couplings are replaced by $g_4$, $g_{2L}$ and $g_{2R}$ for $SU(4)_c$, $SU(2)_L$ and $SU(2)_R$, respectively. The corresponding RGE takes the form
\begin{eqnarray}
	2\pi\frac{{\rm d}\alpha_i}{{\rm d}t}=b_i^{\rm PS}\alpha_i^2\,.
	\label{eq:PSgaugeRGE}
\end{eqnarray}
According to Table.~\ref{tab:particle_contents}, we include only the zero-mode Higgs fields that decomposed from $H_{\bf 10},H_{\bf 120}$, $H_{\bf 16}$ of $SO(10)$ gauge group. Namely, the Higgs contents include $h_1\subset H_{\bf 10}$, $h_1',h_{15}\subset H_{\bf 120}$, $h_{\bar{4}}\subset H_{\bf 16}$. In this case, the $\beta$-coefficients are derived as:
\begin{eqnarray}
(b_4^{\rm PS},b_{2L}^{\rm PS},b_{2R}^{\rm PS})=\Big(-5,\;\frac{7}{3},\;3\Big)\,.
\end{eqnarray}
At the PS scale, the gauge couplings of $SU(4)_c \times SU(2)_R$ are matched to those of $SU(3)_c \times U(1)_Y$ via
\begin{align}
\alpha_{4}^{-1}(M_{\rm PS})=&\; \alpha_{3}^{-1}(M_{\rm PS}) \,, \nonumber\\
\frac{3}{5}\alpha_{2 R}^{-1}(M_{\rm PS})+\frac{2}{5}\alpha_{4}^{-1}(M_{\rm PS})=&\; \alpha_{1}^{-1}(M_{\rm PS}) \,.
\end{align}
For the energy scale  above the compactification scale $M_{\rm KK}$, the RGEs of gauge couplings $g_4$, $g_{2L}$ and $g_{2R}$ are modified to include Kaluza–Klein (KK) contributions, and take the form
\begin{eqnarray}
	2\pi\frac{{\rm d}\alpha_{i}}{{\rm d}t}=b_i^{\rm PS}\alpha_{i}^2+(S(t)-1)b_{\bf10}\alpha_{i}^2\,.
	\label{eq:SO10gaugeRGEED}
\end{eqnarray}
Compared with Eq.~\eqref{eq:RGE_gauge1}, we have added the last term on the right hand side of the above equation to acount for the contributions of Kaluza-Klein (KK) states. Specifically, $S(t)$ includes the contributions of the KK states and can be expressed using a continuous approximation:
\begin{align}
		S(t)=\bigg\{\begin{array}{ll}1&\text{for}\;  \mu<1/R \,,\\ \mu R=M_ZRe^t \;\;&\text{for}\; \mu>1/R \,.\end{array}
	\label{eq:S(t)}
\end{align}
Here, $b_{\bf 10}$ denotes the $\beta$-coefficient associated with the full $SO(10)$ gauge group above $M_{\rm KK}$, including the fifth-dimensional contributions:
\begin{eqnarray}
	b_{\bf10} &=& - (\frac{11}{3}-\frac{1}{6}) C_2(SO(10)) + \frac{4}{3} \sum_F T(F_i) + \frac{1}{6} \sum_S T(S_i)\,, 
	\label{eq:b_10}
\end{eqnarray}
where $\frac{1}{6}C_2(SO(10))$ refers to the contribution from the fifth dimension of the $SO(10)$ gauge bosons, which appear as a real scalar in 4-dimensional spacetime. 

When the energy scale $\mu$ is above the compactification scale $M_{\rm KK}$, the cumulative contribution from Kaluza–Klein (KK) modes must be taken into account. In this regime, both gauge and Yukawa couplings can be recast into effective 't Hooft couplings that incorporate the KK excitations, defined as:
\begin{eqnarray}
	\tilde{\alpha}_i(t)=\alpha_i(t)S(t) \,.
	\label{eq:'t_Hooft_coupling}
\end{eqnarray}
Eq.~\eqref{eq:SO10gaugeRGEED} can then be rewritten in terms of $\tilde{\alpha}$ at high energies. In the high energy limit, it can be approximated as:
\begin{eqnarray}
	2\pi\frac{{\rm d}\tilde{\alpha}_{i}}{{\rm d}t}=2\pi\tilde{\alpha}_{i}+b_{\bf10}\tilde{\alpha}_{i}^2\,.
	\label{eq:SO10gaugeRGElargeED}
\end{eqnarray}
We can solve this differential equation analytically and obtain:
\begin{eqnarray}
	\tilde{\alpha}_{i}=\frac{2\pi}{e^{-t+c_i}-b_{\bf10}}\,,
	\label{eq:analy_solve}
\end{eqnarray}
where $c_i$ (for $i = 4, 2L, 2R$) are constants determined by the value of $\tilde{\alpha}_{4}(M_{\rm KK})$, $\tilde{\alpha}_{2L}(M_{\rm KK})$, and $\tilde{\alpha}_{2R}(M_{\rm KK})$. Regardless of the values of $c_i$, $\tilde{\alpha}_{i}$ will asymptotically approach a common value as the energy scale $\mu$ towards the UV limit:
\begin{eqnarray}
    \tilde{\alpha}_{4}, \tilde{\alpha}_{2L}, \tilde{\alpha}_{2R} \xrightarrow{\text{UV}} \tilde{\alpha}_{\bf10}^{\rm UV} = -\frac{2\pi}{b_{\bf10}}\,,
	\label{eq:fixed_point}
\end{eqnarray}
provided that $b_{\bf10}<0$. In our setup, with a complex Higgs $\bf 10$, a real $\bf 120$, and a chiral-representation Higgs $\bf 16$, $b_{\bf10} = -\frac{19}{3}$, leading to $\tilde{\alpha}_{\bf10}^{\rm UV}=6\pi/19$. This illustrates the concept of asymptotic unification, where all gauge couplings are unified asymptotically at a non-trivial UV fixed point in the UV limit. 

\begin{figure}[t] 
	\begin{center} 
		\includegraphics[width=0.6\textwidth]{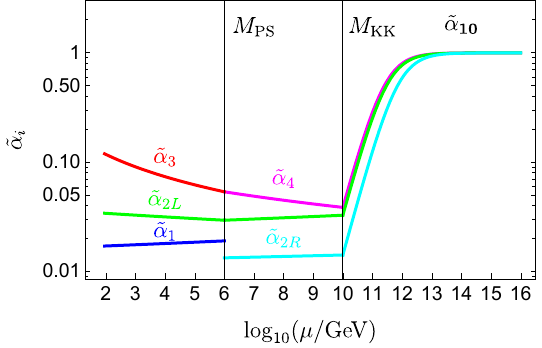}
		\caption{Running of the gauge couplings with $M_{\rm PS}=10^6$ GeV and $M_{\rm KK}=10^{10}$ GeV. Here $\tilde{\alpha}_{\bf10}^{\rm UV} = 6\pi/19$.}
		\label{fig:gauge_running}
	\end{center}
\end{figure}

Naively, this fixed point value may seem to be non-perturbative and thus higher loop contribution may destabilize this fixed point value. However, to properly analyze whether the couplings are perturbative or not, it requires to consider the extra-dimensional loop factor.
The loop factor in generic dimension $d$ can be written as
\begin{equation}
\frac{\Omega(d)}{\left(2\pi\right)^d}\bigg\vert_{d=5}=\frac{8\pi^2}{3\left(2\pi\right)^5}\,,
\end{equation}
where $\Omega(d)$ is the solid angle of a sphere in $d$-dimensional Euclidean space and we have used $\Omega(5)=8\pi^2/3$. Thus, our gauge fixed point value with this new loop factor counting is:
\begin{equation}
\frac{\Omega(d)}{\left(2\pi\right)^d}4\pi\tilde{\alpha}\bigg\vert_{d=5}=\frac{2}{19\pi}\sim\,0.033\ll1.
\end{equation}
Thus, our gauge coupling fixed point value in 5D is well under perturbative control.
Numerically, we show in Fig.~\ref{fig:gauge_running} the running of all gauge couplings from the EW scale to 4D Planck scale\footnote{Once the compactification scale is determined, according to the matching relation between 4D Planck scale and 5D Planck scale for a single extra dimension $M_{\rm Pl}^2=M_*^3(2\pi r)=\frac{2\pi M_*^3}{M_{\rm KK}}$ \cite{Csaki:2004ay}, we can derive the 5D Planck scale $M_*=6.12\times 10^{15}$ GeV directly.}. Without loss of generality, we set $M_{\rm PS}=10^6$ GeV and $M_{\rm KK}=10^{10}$ GeV. The former refers to the lower bound constrained by rare meson decay processes \cite{Valencia:1994cj,Volkas:1995yn}. As shown in the figure, the behavior of the running changes drastically above the compactification scale $M_{\rm KK}$ due to power-law effects from KK excitations, in contrast to the logarithmic running below $M_{\rm KK}$ typical of asymptotically free theories.  The 5D gauge couplings rapidly converge toward the asymptotically safe fixed point $\tilde{\alpha}_{\bf10}^{\rm UV}$ of the 5D $SO(10)$ as shown in Eq.~\eqref{eq:fixed_point}, while the 4D gauge couplings continue to decrease toward zero. This behaviour is consistent with the results in Refs. \cite{Cacciapaglia:2020qky,Khojali:2022gcq}, although our setup includes an additional intermediate gauge symmetry breaking scale below the compactification scale.

\begin{table}[t]
	\begin{center}
		\begin{tabular}{ c c c c c c}
			\hline \hline \\[-4mm]
			Higgs contents & $\tilde{\alpha}_{\bf10}^{\rm UV}$ &  \\[1mm] \hline\\[-4mm]
			
			$({\bf 10}_{\rm r}, {\bf 120}_{\rm r}, {\bf 16})$ & $\displaystyle\frac{4\pi}{13}$ & \\[2mm]
			
			$({\bf 10}_{\rm c}, {\bf 120}_{\rm r}, {\bf 16})$ & $\displaystyle \frac{6\pi}{19}$ & \\[2mm]

			$({\bf 10}_{\rm r}, {\bf 120}_{\rm c}, {\bf 16})$ & $\displaystyle \frac{12\pi}{11}$ & \\[2mm]

			$({\bf 10}_{\rm c}, {\bf 120}_{\rm c}, {\bf 16})$ & $\displaystyle\frac{6\pi}{5}$ & \\[2mm]
									
			$({\bf 10}_{\rm r}, {\bf 120}_{\rm r}, {\bf 45}_{\rm r}, {\bf 16})$ & $\displaystyle\frac{12\pi}{31}$ & \\[2mm]
			
			$({\bf 10}_{\rm r}, {\bf 120}_{\rm r}, {\bf 45}_{\rm r}, {\bf 16})$ & $\displaystyle\frac{12\pi}{31}$ & \\[2mm]

			$({\bf 10}_{\rm c}, {\bf 120}_{\rm r}, {\bf 45}_{\rm r}, {\bf 16})$ & $\displaystyle\frac{2\pi}{5}$ & \\[2mm]
						 
			$({\bf 10}_{\rm r}, {\bf 120}_{\rm r}, {\bf 54}_{\rm r}, {\bf 16})$ & $\displaystyle\frac{4\pi}{9}$ & \\[2mm]

			$({\bf 10}_{\rm r}, \overline{\bf 126})$ & $12\pi$  \\[2mm]

			$({\bf 10}_{\rm c}, \overline{\bf 126})$ & $\infty$  \\[2mm]
			\hline
		\end{tabular}
	\end{center}
	\caption{Potential UV fixed points of the gauge coupling vs Higgs contents of $SO(10)$. The flavour number has been fixed at 3 to match with the SM. The subscript ``r'' or ``c'' of a real representation means its components are assumed to be either real or complex, respectively.
	\label{tab:fixed_points}}
\end{table}

A negative $\beta$-function coefficient $b_{\bf10}$ is crucial to achieve the asymptotic safety of the GUT gauge coupling. This requirement motivates our choice of Higgs representations---$H_{\bf10}$, $H_{\bf120}$ and $H_{\bf16}$---in contrast to the conventional set $H_{\bf10}$, $H_{\bf120}$ and $H_{\overline{\bf126}}$. To make this point explicit, we present below the correlations between $b_{\bf10}$ and multiplicities of fermions and Higgs fields:
\begin{eqnarray}
	b_{\bf10} = -28 + \frac{16}{3} n_{\Psi_{\bf16},{\rm c}} &+&\frac{1}{6} n_{H_{\bf10},{\rm r}}+\frac{14}{3} n_{H_{\bf120},{\rm r}}+\frac{4}{3} n_{H_{\bf45},{\rm r}}+2 n_{H_{\bf54},{\rm r}} \nonumber\\
	&+&\frac{35}{3} n_{H_{\overline{\bf126}},{\rm c}}+\frac{2}{3}n_{H_{\bf16},{\rm c}} + \cdots\,, 
	\label{eq:b_10_v2}
\end{eqnarray}
Here $n_f$ denotes the copy of fermion pairs $(\Psi_{\bf16},\Psi_{\overline{\bf16}})$; $n_{H_{\bf10},{\rm r}}$, $n_{H_{\bf120},{\rm r}}$, $n_{H_{\bf45},{\rm r}}$ and $n_{H_{\bf54},{\rm r}}$ denote the copies of Higgs fields in real representations $\bf10$, $\bf120$, $\bf45$ and $\bf54$, respectively;  $n_{H_{\bf126},{\rm c}}$, $n_{H_{\bf16},{\rm c}}$ denote the copies of Higgs fields in complex representations $\overline{\bf126}$ and $\bf16$, respectively. The ellipsis indicates possible contributions from any additional particles. In our model, we obtain $b_{\bf10}=-\frac{19}{3}$ by choosing $n_{\Psi_{\bf16}}=3$ to account for three generations of SM fermions, $n_{H_{\bf10},{\rm r}}=2$ for a complex $H_{\bf10}$, $n_{H_{\bf120},{\rm r}}=1$ for real $H_{\bf120}$, and $n_{H_{\bf16},{\rm c}}=1$ for a chiral Higgs. No other contributions are assumed, as summarized in Table~\ref{tab:particle_contents}. Below, we fix $n_{\Psi_{\bf16}} = 3$ to match the three flavours in the SM and explore how $b_{\bf10}$ and the UV fixed point $\tilde{\alpha}_{\rm UV}$ depend on the variations of Higgs multiplets.

We begin by explaining why the inclusion of $H_{\overline{\bf126}}$ is incompatible with the goal of achieving an asymptotically safe GUT. The  field $H_{\overline{\bf126}}$ belongs to a 126-dimensional complex representation.\footnote{We adopt a left-right convention, which differs from the widely used right-left convention in SUSY GUTs. In our convention, the Yukawa coupling involving $H_{\overline{\bf126}}$ takes the form $\overline{\Psi_{\bf16}} H_{\overline{\bf126}}^*\Psi_{\overline{\bf16}} + {\rm h.c.}$.} It contributes to the $\beta$-function coefficient with a large and positive factor $35/3$. Including a single $H_{\overline{\bf126}}$ yields $b_{\bf10} = -1/3$ and a corresponding UV fixed point value of $\tilde{\alpha}_{\bf10}^{\rm UV}  = 6 \pi$. Since a single $H_{\overline{\bf126}}$ is insufficient to generate realistic mass splittings for quarks and leptons, one might consider supplementing it with a real $H_{\bf10}$. In this case, the coefficient becomes $b_{\bf10}=-1/6$, pushing the UV fixed point to $\tilde{\alpha}_{\bf10}^{\rm UV}  = 12 \pi$. Replacing the real $H_{\bf10}$ with a complex one further increases $b_{\bf10}$ to zero, leading to a divergent UV coupling $\tilde{\alpha}_{\bf10}^{\rm UV}  \to \infty$. Adding more Higgs multiplets only increases $b_{\bf10}$, driving it positive and thereby spoiling the possibility of asymptotic safety and thus fail to support asymptotic gauge unification.

In this work, we adopt a minimal Higgs sector consisting of ($H_{\bf10}$, $H_{\bf120}$ and  $H_{\bf16}$). Nevertheless, this choice is not unique for realizing the asymptotic safety of the gauge coupling. While the use of $H_{\overline{\bf126}}$ is excluded, several viable alternatives remain. For instance, one may consider varying the real or complex nature of the $H_{\bf10}$ and $H_{\bf120}$ fields, or introducing additional representations such as ${\bf45}$ or ${\bf54}$, which are commonly associated with intermediate gauge symmetries. A selection of possible UV fixed points corresponding to different Higgs configurations is presented in Table~\ref{tab:fixed_points}. Among these, the Higgs content adopted in this work—$({\bf10}{\rm c}, {\bf120}{\rm r}, {\bf16})$—represents the most economical and phenomenologically viable option, capable of accommodating realistic mass spectra for quarks, charged leptons, and neutrinos.

\section{Running of the Yukawa couplings} \label{sec:4}

In this section, we analyze the UV behaviour of the Yukawa couplings. Given the small masses of the first- and second-generation fermions, we restrict our analysis to the third-generation Yukawa couplings for simplicity. The running of the fundamental Yukawa couplings with respect to the energy scale $\mu$, from the EW scale to the deep UV is sketched as follows
\begin{align}
  \begin{array}{cccccccccccc}
   \mu: &&  M_Z & \mathrel{\tikz\draw[->, thick] (0,0) -- (15mm,0);} & M_{\rm PS} & \mathrel{\tikz\draw[->, thick] (0,0) -- (15mm,0);} & M_{\rm KK} & \mathrel{\tikz\draw[->, thick] (0,0) -- (15mm,0);} & {\rm UV} \\
   {\rm Yukawas}:&&& y_t,y_b,y_\tau && y_1, y_1', y_{15} && y_1, y_1', y_{15} & \quad y_{\bf10}, y_{\bf120} \,.
   \end{array}
\end{align} 
The explicit one-loop Yukawa RGEs involving full three families are listed in Appendices~\ref{app:2} and \ref{app:3}. 
In this analysis, we neglect the running of $y_{\bf16}$ and, consequently, the couplings $y_4$, which originate from the interactions with the spinor Higgs $H_{\bf16}$ and its PS decomposed components $h_{\bar{4}}$, respectively. These couplings primarily serve to explain the smallness of light neutrino masses. Their contributions to the other Yukawas' RGEs can be safely neglected under the assumption that they are much smaller than the rest.  For completeness, the RGEs including $y_4$ and $y_{\bf16}$ are also provided in Appendices~\ref{app:2} and \ref{app:3}.

\subsection{RGEs of Yukawa couplings}

We begin our analysis by examining the RG evolution of the Yukawa couplings from the EW scale to the deep UV regime. 

From the EW scale to the PS scale, the Yukawa couplings of $t$, $b$ and $\tau$ under the SM gauge symmetries are governed by the RGEs below:
\begin{align}
		&2\pi \frac{{\rm d}\alpha_{t}}{{\rm d}t} = \Big[\frac{9}{2} \alpha_t+\frac{3}{2}\alpha_b+\alpha_\tau-\frac{9}{4}\alpha_{2L} -\frac{17}{20}\alpha_1 -8 \alpha_3 \Big] \alpha_t \,, \nonumber\\
		&2\pi \frac{{d}\alpha_{b}}{{\rm d}t} = \Big[\frac{3}{2} \alpha_t+\frac{9}{2}\alpha_b+\alpha_\tau -\frac{9}{4}\alpha_{2L} -\frac{1}{4}\alpha_1-8 \alpha_3 \Big] \alpha_b \,, \nonumber\\
		&2\pi \frac{{\rm d}\alpha_\tau}{{\rm d}t} = \Big[ 3 \alpha_t+3 \alpha_b+\frac{5}{2} \alpha_\tau -\frac{9}{4}\alpha_{2L} -\frac{9}{4}\alpha_1 \Big] \alpha_\tau \,,
		\label{eq:Yukawa_RGE_SM}
\end{align}
where $\alpha_f = y_f^2/(4\pi)$. 

From $M_{\rm PS}$ to the compactification scale $M_{\rm KK}$, the fundamental Yukawa couplings change from $y_t$, $y_b$, $y_\tau$ to $y_1$, $y_1'$, $y_{15}$. The matching conditions at the PS scale are provided in Eq.~\eqref{eq:Yukawa_PS}. 
The RGEs for the redefined couplings $\alpha_{y{\rm r}} = y_{y{\rm r}}^2/(4\pi)$, with ${\rm r}=1,1',15$, are given by:
\begin{align}
		2\pi\frac{{\rm d}\alpha_{y1}}{{\rm d}t}=&\Big[ 6 \alpha_{y1} + 4\alpha_{y1'}-\frac{45}{4} \alpha_4 -\frac{9}{4}( \alpha_{2L} + \alpha_{2R} )\Big] \alpha_{y1} \,, \nonumber\\
		2\pi\frac{{\rm d}\alpha_{y1'}}{{\rm d}t}=&\Big[ 2\alpha_{y1} + 8\alpha_{y1'}-\frac{45}{4} \alpha_4 -\frac{9}{4}( \alpha_{2L} + \alpha_{2R} )\Big] \alpha_{y1'} \,, \nonumber\\
		2\pi\frac{{\rm d}\alpha_{y15}}{{\rm d}t}=&\Big[ 8\alpha_{y15} + 2 \alpha_{y1} - \frac{45}{4} \alpha_4 -\frac{9}{4}( \alpha_{2L} + \alpha_{2R} ) \Big] \alpha_{y15} \,. 
	\label{eq:Yukawa_RGE_PS}
\end{align}

For the energy scale $\mu$ above the compactification scale $M_{\rm KK}$, we should consider full particle contents in $SO(10)$. Full RGEs of any Yukawa couplings $y_{\rm r}$ are then replaced by:
\begin{align}
		2\pi\frac{{\rm d}\alpha_{y\rm r}}{{\rm d}t}=&2\pi\frac{{\rm d}\alpha_{y\rm r}}{{\rm d}t}\Big|_{\rm PS} + (S(t)-1)\, 2\pi\frac{{\rm d}\alpha_{y\rm r}}{{\rm d}t}\Big|_{\rm KK} \,,
	\label{eq:Yukawa_RGE_KK}
\end{align}
where the first term on the RHS represents contributions in Eq.~\eqref{eq:Yukawa_RGE_PS}, and the second term, proportional to $(S(t) - 1)$, accounts for the d.o.f.~of KK states above the compactification scale. The explicit KK-induced contributions to the beta functions are given by: 
\begin{align}
		2\pi\frac{{\rm d}\alpha_{y1}}{{\rm d}t}\Big|_{\rm KK} =&\Big[10\alpha_{y1} +\frac{3}{2}\alpha_{y6} + 4\alpha_{y1'} - \frac{5}{2} \alpha_{y10} +\frac{9}{4}\alpha_{y6'} - \frac{81}{8} \alpha_{4} \nonumber\\
		&- \frac{45}{8} (\alpha_{2L} + \alpha_{2R})\Big] \alpha_{y1} \,, \nonumber\\
		2\pi\frac{{\rm d}\alpha_{y1'}}{{\rm d}t}\Big|_{\rm KK} =&\Big[2\alpha_{y1} +\frac{3}{2}\alpha_{y6} + 12\alpha_{y1'} +\frac{15}{2}\alpha_{y10} +\frac{9}{4}\alpha_{y6'} - \frac{129}{8} \alpha_{4} \nonumber\\
		& - \frac{45}{8} (\alpha_{2L} + \alpha_{2R})\Big] \alpha_{y1'} \,, \nonumber\\
		2\pi\frac{{\rm d}\alpha_{y15}}{{\rm d}t}\Big|_{\rm KK} =&\Big[2\alpha_{y1} + \frac{3}{2} \alpha_{y6} + 9\alpha_{y15} + \frac{3}{2} \alpha_{y10} + \frac{9}{4}\alpha_{y6'}- \frac{129}{8} \alpha_4 \nonumber\\
		&- \frac{45}{8}( \alpha_{2L} + \alpha_{2R} ) \Big] \alpha_{y15} \,. 
	\label{eq:Yukawa_RGE_KK2}
\end{align}
Here, couplings $\alpha_{y6}$, $\alpha_{y6'}$ and $\alpha_{y10}$ are the new KK states contributions via the Yukawa couplings:
\begin{align}
&y_6 ( \overline{\psi_L} H_6 \Psi_L^c +  \overline{\Psi_R^c} H_6 \psi_R ) + y_6' ( \overline{\psi_L} H_{6L} \Psi_L^c +  \overline{\Psi_R^c} H_{6R} \psi_R ) \nonumber\\
& + y_{10} ( \overline{\psi_L} H_{10} \Psi_L^c +  \overline{\Psi_R^c} H_{\overline{10}} \psi_R ) + {\rm h.c.} \,,
\end{align} 
which appear only above the KK scale. We refer the reader to Appendix~\ref{app:4} for more details.
Similar to the treatment of gauge couplings, it is convenient to introduce the effective 't Hooft coupling for each Yukawa coupling to describe its RG running above the KK scale:
\begin{eqnarray}
\tilde{\alpha}_{y{\rm r}}(t)=\alpha_{y{\rm r}}(t) S(t) \,.
\end{eqnarray} 
The RGEs in Eq.~\eqref{eq:Yukawa_RGE_KK} can then be reformulated as differential equations for the ’t Hooft couplings, which will not be repeated here. 

\begin{figure}[t] 
	\begin{center} 
		\includegraphics[width=0.49\textwidth]{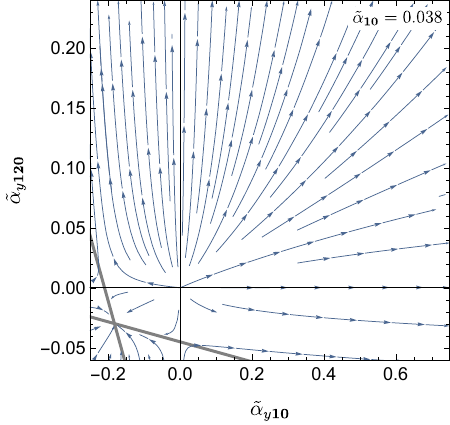}
		\includegraphics[width=0.49\textwidth]{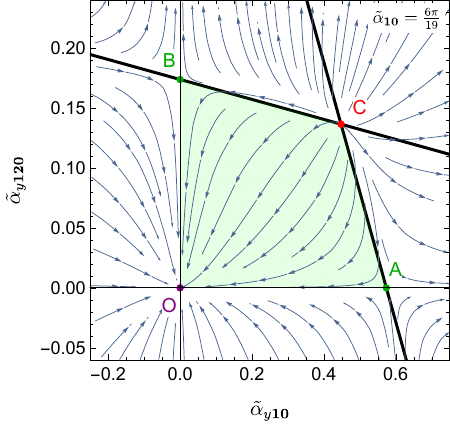}
		\caption{Stream plot of the Yukawa couplings $\tilde{\alpha}_{y\bf10}, \tilde{\alpha}_{y\bf120}$ in the 5D $SO(10)$ GUT. The black lines denote the separtirx lines connecting two pairs of fixed points, $AC$ and $BC$, dividing the phase diagram into distinct regions. The green region represents the asymptotically free phase, where the couplings flow toward the Gaussian fixed point $O$ in the UV. The gauge coupling $\tilde{\alpha}_{\bf10}$ is held fixed in each plot: $\tilde{\alpha}_{\bf10} = 0.038$ in the left panel and $\tilde{\alpha}_{\bf10}=\tilde{\alpha}_{\bf10}^{\rm UV} = 6\pi/19$ in the right panel. In the left panel, apart from a non-physical region where $\tilde{\alpha}_{y\bf120}$ becomes negative and the flow appears asymptotically safe, no asymptotically free region is observed, indicating that a sufficiently large gauge coupling is essential for realizing asymptotic freedom of the Yukawa sector.}
		\label{fig:potential_free}
	\end{center}
\end{figure}

For energy scales well above the KK scale, if asymptotic unification is realized, the PS Yukawa couplings are expected to approach the fundamental $SO(10)$ Yukawa couplings. 
Since all PS Yukawa couplings have been unified into those of $SO(10)$, it is convenient to replace the 't Hooft couplings in the PS model by their $SO(10)$ counterparts. The RGEs for the relevant 't Hooft couplings are then given by:
\begin{align}
		2\pi\frac{{\rm d}\tilde{\alpha}_{y\bf10}}{{\rm d}t}=& \Big[ 2\pi+26\tilde{\alpha}_{y\bf10}+24\tilde{\alpha}_{y\bf120} -\frac{171}{8}\tilde{\alpha}_{\bf10} \Big] \tilde{\alpha}_{y\bf10} \,,\nonumber\\
		2\pi\frac{{\rm d}\tilde{\alpha}_{y\bf120}}{{\rm d}t}=& \Big[ 2\pi+10\tilde{\alpha}_{y\bf10} +120\tilde{\alpha}_{y\bf120} -\frac{219}{8}\tilde{\alpha}_{\bf10} \Big] \tilde{\alpha}_{y\bf120} \,.
	\label{eq:Yukawa_RGE_SO10_ED}
\end{align}
These equations are obtained in the  $SO(10)$ framework and consistent with Eq.~\eqref{eq:Yukawa_RGE_tHooft} once the matching condition in Eq.~\eqref{eq:matching_SO10_PS} in the appendix is considered. 

We investigate the intriguing possibility that the Yukawa couplings flow towards a fixed point. Similar to gauge couplings, we examine whether the $\beta$ function exhibits zeros at high energy. On the right panel of Fig.~\ref{fig:potential_free}, we set the gauge couplings at their UV fixed point value, $\tilde{\alpha}_{\bf10}={6\pi}/{19}$. The three non-trivial fixed points solutions $A$, $B$, $C$ derived from Eq.~\eqref{eq:Yukawa_RGE_SO10_ED}, together with the trivial Gaussian fixed point $O$, are given below:
\begin{equation}
\begin{array}{ccccc}
	(\tilde{\alpha}_{y\bf10}, \tilde{\alpha}_{y\bf120} )  = & (0,0), & (\frac{19\pi}{104}, 0), & (0, \frac{101\pi}{1824}), &
	(\frac{65\pi}{456}, \frac{119\pi}{2736}). \\
	\text{Point} & O & A & B & C \\
\end{array}
	\label{eq:UVfixed}
\end{equation}
Separatrix lines are constructed by connecting the fixed points $AC$ and $BC$ in the phase diagram, shown in black in the stream plot on the right panel of Fig.~\ref{fig:potential_free}. These separatrix lines clearly divide the total phase diagram into two regions. In the green region of Fig.~\ref{fig:potential_free}, with the convention of RG flow direction from IR to UV, the trivial Gaussian fixed point $O$ acts as a fully attractive UV fixed point while $C$ acts as a fully repulsive IR fixed point. Thus, the green region represents the asymptotically free phase. There are two single solutions running along the separatrix (i.e.~from $C$ to $A$ and from $C$ to $B$) that can be asymptotically safe, but they are highly fine-tuned due to the mixed characteristics of the fixed points $A$ and $B$ (which exhibit both ultraviolet and infrared properties), and we ignore these two cases in our analysis.

The asymptotically free region shown above corresponds to a specific 2D slice of the full 3D RG flow of $\left(\tilde{\alpha}_{\bf10},\,\tilde{\alpha}_{y\bf10},\,\tilde{\alpha}_{y\bf120}\right)$, obtained by fixing $\tilde{\alpha}_{\bf10}$ at its UV fixed point.
In general, varying the value of $\tilde{\alpha}_{\bf10}$ changes the corresponding 2D slice and, consequently, alters the asymptotically free region of Yukawa coupling. In realistic RG running, $\tilde{\alpha}_{\bf10}$ evolves from percent-level value at the KK scale to its fixed point value in the UV limit as shown in Fig.~\ref{fig:gauge_running}. If a smaller value of $\tilde{\alpha}_{\bf10}$ is adopted along the RG trajectory in Fig.~\ref{fig:gauge_running}, the asymptotically free region progressively shrinks and can eventually vanish, as clearly shown on the left panel of Fig.~\ref{fig:potential_free}, where $\tilde{\alpha}_{\bf10} = 0.038$. In this case, no physical asymptotically free solution remains, except for a non-physical asymptotically safe region where $\tilde{\alpha}_{y\bf120}$ becomes negative—underscoring the necessity of a sufficiently large gauge coupling to achieve Yukawa asymptotic freedom.
From this perspective, the asymptotically free region shown in the right panel of Fig.~\ref{fig:potential_free} represents the most ideal scenario, where the region is maximized. A careful analysis of the Yukawa RG trajectories over the full energy range is therefore essential to confirm whether the theory indeed exhibits asymptotic freedom. We will revisit this point in section~\ref{sec:4.3} through a detailed benchmark study.

We now turn to the matching between Yukawa couplings in the PS gauge symmetry and their counterparts in $SO(10)$. Naively, following the concept of asymptotic unification, one may expect the matching condition to be
\begin{eqnarray}
\begin{array}{ccc}
	\displaystyle
		\frac12 \alpha_{y1},\, \frac14 \alpha_{y6} &\to & \alpha_{y\bf10}\,, \\[2mm]
	\displaystyle
		\frac12 \alpha_{y1'},\,\frac18 \alpha_{y15},\, \frac18 \alpha_{y10}, \frac{1}{16} \alpha_{y6'} &\to&   \alpha_{y\bf120}\,,
\end{array} \quad \text{in the UV limit,} 
	\label{eq:Yukawa_UV}
\end{eqnarray} 
where $\alpha_{y\bf10} = y_{\bf10}^2 / (4\pi)$, the prefactors follow from group-theoretical considerations in Eq.~\eqref{eq:matching_SO10_PS}. However, a careful analysis shows that if two Yukawa couplings, such as $\frac{1}{2} \alpha_{y1'}$ and $\frac18 \alpha_{y15}$, do not coincide at the KK scale, it is hard for them to converge to the same value in the deep UV regime. This is because they share the same gauge contribution which dominate in the UV limit. Their ratio will thus remain constant along the RG flow. To avoid this situation, we impose the ``exact unification condition" for the Yukawa couplings, assuming that all Yukawa couplings have already been matched to their $SO(10)$ values at the KK scale:
\begin{eqnarray}
\begin{array}{ccc}
	\displaystyle
                 \frac12 \alpha_{y1},\, \frac14 \alpha_{y6} &=& \alpha_{y\bf10} \,, \\[2mm]
	\displaystyle
		\frac12 \alpha_{y1'},\,\frac18 \alpha_{y15},\, \frac18 \alpha_{y10}, \frac{1}{16} \alpha_{y6'} \hspace{3mm} &=&  \alpha_{y\bf120} \,,
\end{array} \quad \text{at }\mu = M_{\rm KK} \,.  
\label{eq:explicit_condition}
\end{eqnarray} 
This condition imposes a stronger constraint than the asymptotic unification of gauge couplings, as it requires exact matching at the compactification scale rather than convergence in the UV limit. Under this condition, the RGEs for the relevant 't Hooft couplings $\tilde{\alpha}_{y\bf10}$, $\tilde{\alpha}_{y\bf120}$ are then explicitly written as
\begin{align}
		2\pi\frac{{\rm d}\tilde{\alpha}_{y\bf10}}{{\rm d}t}=& \Big[2\pi+26\tilde{\alpha}_{y\bf10}+24\tilde{\alpha}_{y\bf120} - \frac{81}{8} \tilde{\alpha}_{4} - \frac{45}{8} (\tilde{\alpha}_{2L} + \tilde{\alpha}_{2R})\Big] \tilde{\alpha}_{y\bf10} \,, \nonumber\\
		2\pi\frac{{\rm d}\tilde{\alpha}_{y\bf120}}{{\rm d}t}=& \Big[ 2\pi+10\tilde{\alpha}_{y\bf10} +120\tilde{\alpha}_{y\bf120} - \frac{129}{8} \tilde{\alpha}_{4} - \frac{45}{8} (\tilde{\alpha}_{2L} + \tilde{\alpha}_{2R})\Big] \tilde{\alpha}_{y\bf120} \,. 
	\label{eq:Yukawa_RGE_tHooft}
\end{align}

\subsection{Asymptotic behaviour of Yukawa couplings}

We examine the RG running of the Yukawa couplings in the asymptotic $SO(10)$ GUT with an extra dimension.  
The procedure for the numerical running is outlined as follows. 
To satisfy the matching condition between the PS and $SO(10)$ Yukawa couplings at the compactification scale, as specified in Eq.~\eqref{eq:explicit_condition}, we initiate the scan by setting 
\begin{align}
y_{\bf 10}(M_{\rm KK}) \in (10^{-2},\sqrt{2\pi}) \,,\quad
y_{\bf 120}(M_{\rm KK})\in (10^{-2},\sqrt{\pi/2}) \,,
\end{align} 
which ensures $\alpha_{y{\bf 1}}< 1, \alpha_{y{\bf 15}}< 1$ i.e.~remain in the perturbative regime. The values of $y_1(M_{\rm KK}),y'_{1}(M_{\rm KK}),y_{15}(M_{\rm KK})$ are then determined from the inputs $y_{\bf 10}(M_{\rm KK})$ and $y_{\bf 120}(M_{\rm KK})$ via the matching condition in Eq.~\eqref{eq:explicit_condition}. These three Yukawa couplings are subsequently evolve from the compactification scale to the PS breaking scale $M_{\rm PS}$ using the RGEs in Eq.~\eqref{eq:Yukawa_RGE_PS}, yielding $y_1(M_{\rm PS}),y'_{1}(M_{\rm PS}),y_{15}(M_{\rm PS})$. 

On the other hand, at the EW scale, the following Yukawa couplings in the modified minimal subtraction $(\overline{\rm MS})$ scheme\cite{Huang:2020hdv} are taken as initial values for Eq.~\eqref{eq:Yukawa_RGE_SM} to evolve the couplings from the EW scale to $M_{\rm PS}$
\begin{eqnarray}
	\{y_t, y_b,y_\tau\}=\{0.97,0.016,0.010\} \; \text{at} \; \mu = M_Z \,.
\end{eqnarray}
Given these values, we evolve the Yukawa couplings via the SM RGEs in Eq.~\eqref{eq:Yukawa_RGE_SM}, and obtain the values of $y_t,y_b,y_\tau$ at the PS breaking scale. The matching condition at this scale, presented in Eq.~\eqref{eq:Yukawa_PS}, relates these quantities to the PS Yukawa couplings $y_1$, $y_1'$ and $y_{15}$, which are themselves obtained by evolving down from the KK scale. For consistency, the two sets of Yukawa couplings—those from the KK scale and those from the EW scale—must satisfy the matching condition simultaneously, which in turn uniquely determines the values of the Higgs VEV coefficients $c_{\bf10}^u$, $c_{\bf10}^d$, $c_{\bf120}^d$, $c_{\bf120}^{d'}$. These  coefficients are treated as energy independent, and are subject to the normalization constraint in Eq.~\eqref{eq:vev}. Assuming all these mixing parameters are real, the constraint simplifies to:
\begin{align}
(c_{\bf10}^u)^2+(c_{\bf10}^d)^2 + 2(c_{\bf120}^d)^2+ 2(c_{\bf120}^{d'})^2 = 1 \,.
\label{eq:vevr}
\end{align}
To ensure a consistent matching, we scan over two free input parameters: 
\begin{eqnarray}
	\{y_{\bf10} (M_{\rm KK}),\,y_{\bf120} (M_{\rm KK})\} ,\,
\end{eqnarray}
from which the values of all dependent Yukawa couplings and VEV coefficients are determined.

\begin{figure}[t] 
	\begin{center} 
		\includegraphics[height=0.3\textwidth]{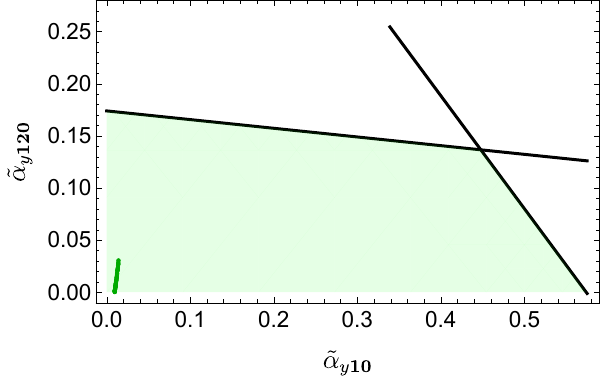}
		\includegraphics[height=0.3\textwidth]{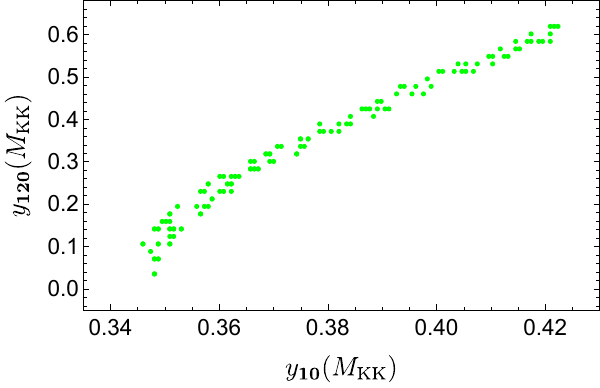}
		\includegraphics[height=0.3\textwidth]{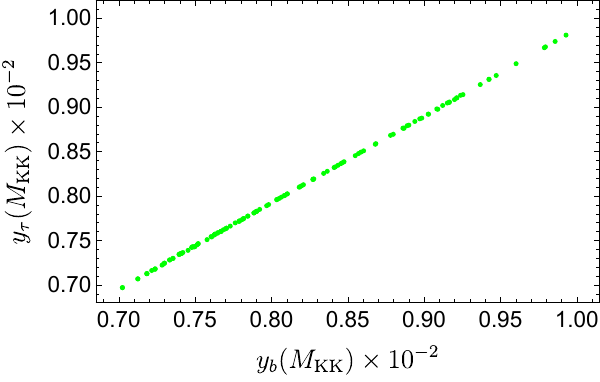}
		\includegraphics[height=0.3\textwidth]{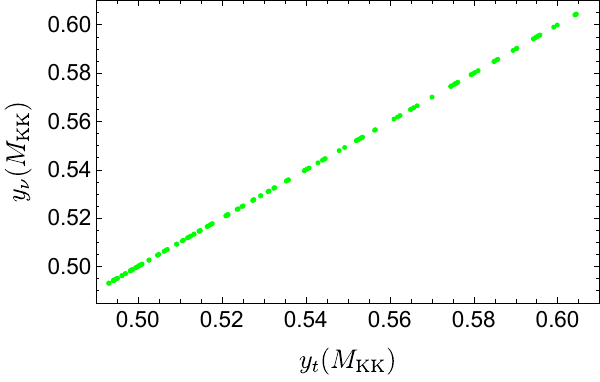}
		\caption{Two-dimensional correlations between theory inputs and outputs at the KK scale. Points in the potential asymptotically free region are labelled by green points, the remaining points out of the potential asymptotically free region are labelled by gray points. $M_{\rm PS}=10^6$ GeV and $M_{\rm KK}=10^{10}$ GeV are determined. The predicted charged fermion masses and Dirac neutrino mass term $m_{\rm D}$ are calculated at the compactification scale $M_{\rm KK}$.}
		\label{fig:scan}
	\end{center}
\end{figure}

Finally, inserting the scanned values of $y_{\bf10}(M_{\rm KK}),y_{\bf120}(M_{\rm KK})$ into Eq.~\eqref{eq:Yukawa_RGE_tHooft}, we obtain the full RG evolution of the Yukawa couplings. At the PS-breaking scale $M_{\rm PS}$, the coupling $y_{15}(M_{\rm PS})=\frac{\sqrt{3}(y_b(M_{\rm PS})-y_\tau(M_{\rm PS}))}{2c_{\bf120}^{d}}$ follows from the matching relation in Eq.~\eqref{eq:Yukawa_PS}. In particular, $y_{15}(M_{\rm PS})$ vanishes if $M_{\rm PS}$ is set at $1.6\times 10^{6} \ {\rm GeV}$, where the $b$-$\tau$ unification is realised. 
The results of the parameter scan are presented in Fig.~\ref{fig:scan}, including the predicted charged fermion masses and the Dirac neutrino mass term $m_{\rm D}$ at the compactification scale. All valid points lie within the asymptotically free region (green points), confirming consistency with the UV behaviour. Among them, the input parameters and low-energy Yukawa couplings span the following ranges:
\begin{align*}
y_{\bf10}(M_{\rm KK}) &\in (0.3460,\, 0.4223)\,, \\
y_{\bf120}(M_{\rm KK}) &\in (0.0353,\, 0.6187)\,, \\
y_b(M_{\rm KK}) &\in (0.0070,\, 0.0099)\,, \\
y_\tau(M_{\rm KK}) &\in (0.0070,\, 0.0098)\,, \\
y_t(M_{\rm KK}) &\in (0.4931,\, 0.6044)\,, \\
y_{\nu}(M_{\rm KK}) &\in (0.4930,\, 0.6043)\,.
\end{align*}
According to the matching condition in Eq.~\eqref{eq:Yukawa_PS}-with $H_{\bf 10}$ taken to be complex and  $H_{\bf 120}$ real-the sizable mass splitting between top and bottom quarks i.e.~$y_t-y_b$ arises from the term $y_1(c_{\bf 10}^u-c_{\bf 10}^d)$. Therefore, $c_{\bf 10}^u$ must of $O(1)$ while $c_{\bf 10}^d$ should be rather small, which is clearly shown in Table.~\ref{tab:benchmark} below.

We now address the role of Yukawa couplings in generating the light neutrino mass. In our RG analysis, the Yukawa coupling $y_{\bf16}$, associated with the neutrino sector, has been neglected. However, as indicated by the effective light neutrino mass matrix in Eq.\eqref{eq:effective_neutrino_matrix}, the resulting light neutrino mass still depends on the Dirac mass $m_{\rm D}$, the coupling $y_{16}$ and the Majorana mass scale $\mu_{\rm M}$ to obtain the light neutrino mass. Thus, to produce the sub-eV neutrino mass, $y_{16}$ must remain sufficiently large to compete with the suppression from $\mu_{\rm M}$. Specifically, the doubly-suppressed term $\frac{m_{\rm D}}{y_{16}v_{\rm S}} \sim \frac{m_t}{y_{16}M_{\rm PS}}$, together with a small Majorana mass $\mu_{\rm M}$, can yield viable light neutrino mass. For instance, choosing $\mu_{\rm M}=1$~keV gives $y_{16}=10^{-2}$. This value is much smaller than the other Yukawa couplings in the theory, thereby justifying the neglect of its contribution in their RG running. 

\subsection{Benchmark point}\label{sec:4.3}

We conclude this section by presenting a benchmark study on the RG running from the EW scale to the deep UV. 
The input values and the resulting predictions for the VEVs, as well as the Yukawa couplings of charged fermions and neutrinos at the benchmark point, are summarized in Table.~\ref{tab:benchmark}. With $y_{\bf16} = 10^{-2}$, a light neutrino mass is found to be $m_\nu = 0.07$~eV. Note that in this benchmark setup, $y_{\bf16} \ll y_{\bf10}, y_{\bf120}$, so its contribution to RGEs of $y_{\bf10}$ and $y_{\bf120}$ is negligible.

\begin{table}[h] 
	\centering
	\begin{tabular}{m{2cm}<{\centering}|m{2.5cm}<{\centering}|m{2.5cm}<{\centering}|m{2.5cm}<{\centering}|m{2.5cm}<{\centering}}
		\hline \hline
		\multirow{6}{*}{Inputs} & $y_{\bf10}(M_{\rm KK})$ & $y_{\bf120}(M_{\rm KK})$ &  & \\
		& 0.348 & 0.035 &  &  \\
		\cline{2-5}
		& $M_{\rm PS}$ & $M_{\rm KK}$ & $\mu_{\rm M}$ & $y_{\bf16}(M_{\rm KK})$ \\
		& $10^6$ GeV & $10^{10}$ GeV & 1 keV & $10^{-2}$ \\
		\hline
		\multirow{6}{*}{Outputs} & $c_{\bf10}^u$ & $c_{\bf10}^d$ & $c_{\bf120}^{d}$ & $c_{\bf120}^{d'}$ \\
		& 0.999 & 0.013 & 0.0004 & 0.0217 \\
		\cline{2-5}
		& $y_{1}(M_{\rm PS})$ & $y'_{1}(M_{\rm PS})$ & $y_{15}(M_{\rm PS})$ &$m_\nu$ \\
		& 0.681 & 0.075 & 0.149 & 0.07 eV \\
		\cline{2-5}
		& $y_b(M_{\rm KK})$ & $y_\tau(M_{\rm KK})$ & $y_t(M_{\rm KK})$ & $y_{\nu}(M_{\rm KK})$  \\
		& 0.0074 & 0.0073 & 0.534 & 0.493 \\
		\hline
		\hline
	\end{tabular}
	\caption{Inputs and predictions of VEVs, Yukawa couplings, charged fermion masses and neutrino masses of one point.}	\label{tab:benchmark}
\end{table}
\begin{figure}[h] 
	\begin{center} 
		\includegraphics[width=0.7\textwidth]{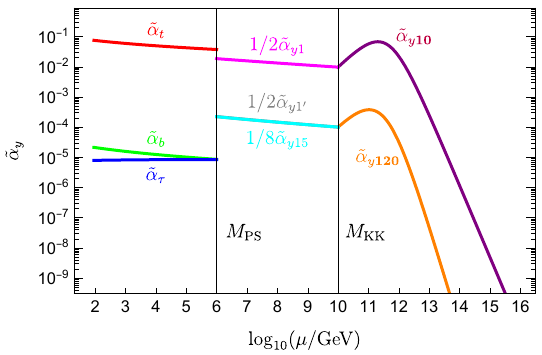}
		\caption{Running of the Yukawa couplings for the benchmark point with $M_{\rm PS}=10^6$ GeV and $M_{\rm KK}=10^{10}$ GeV.}
		\label{fig:Yukawa_running}
	\end{center}
\end{figure}

In Fig.~\ref{fig:Yukawa_running}, we display the running of the Yukawa couplings from the EW scale to the 4D Planck scale. For energy scale above the KK scale, the evolution of $\tilde{\alpha}_{y\bf 10}$ and $\tilde{\alpha}_{y\bf 120}$  is governed by Eq.~\eqref{eq:Yukawa_RGE_tHooft}. 
We observe that $\tilde{\alpha}_{y\bf 10}$ and  $\tilde{\alpha}_{y\bf 120}$ initially increase with the energy scale, especially just above the KK scale. This behaviour arises because the gauge couplings are still small at that scale, and their negative contribution to the $\beta$ functions are insufficient to overcome the positive contribution from Yukawa couplings. As the energy scale increases further, around $\mu \sim 10^2 M_{\rm KK}$ in this benchmark, the gauge couplings approach their fixed point---typically of $O(1)$---such that their contributions to the $\beta$ function become dominant. This causes the Yukawa couplings to decrease with energy scale. At even higher scale, the $\beta$ functions are entirely dominated by the gauge contributions, and the Yukawa couplings begin to exhibit asymptotically free behaviour.
In this UV regime, governed by Eq.~\eqref{eq:Yukawa_RGE_SO10_ED}, the Yukawa couplings obey the following scaling laws:
\begin{align}
    \tilde{\alpha}_{y\bf10} \propto&\, \mu^{-\frac{171}{16\pi} \tilde{\alpha}_{\bf10}^{\rm UV}+1} = \mu^{-19/8} \,, \nonumber\\ 
    \tilde{\alpha}_{y\bf120} \propto&\, \mu^{-\frac{219}{16\pi} \tilde{\alpha}_{\bf10}^{\rm UV}+1} = \mu^{-505/152}  \,.
\end{align} 
Unlike the conventional logarithmic running in four-dimensional theories, the power-law decrease of Yukawa couplings in the UV arises from the higher-dimensional structure of the theory, where the cumulative contributions of KK towers significantly enhance the gauge-induced suppression.

\section{Conclusion} \label{sec:5}

Asymptotic grand unification provides an alternative paragdim for the unification of gauge couplings. Rather than postulating exact unification at a specific high energy scale, it proposes that the couplings asymptotically converge in the deep UV limit. 
In this work, we demonstrate that asymptotic grand unification is realized in a five-dimensional $SO(10)$ model with minimal and realistic field content.   

The model is briefly summarized as follows: 1) A 5D $SO(10)$ GUT is assumed to describe the deep UV regime; 2) it is broken to the Pati-Salam (PS) gauge symmetry, $SU(4)_c \times SU(2)_L \times SU(2)_R$, via boundary conditions (BCs) at the compactification scale, i.e., the Kaluza-Klein (KK) scale; 3) The PS symmetry is further broken to the SM gauge symmetry via the Higgs mechanism, triggered by a spinor Higgs field $H_{\bf16}$ at lower scale. Without loss of generality, we assumed the KK scale to be $10^{10}$~GeV and the PS breaking scale to be around $10^6$~GeV. 
The particle content is arranged as follows: 
Two bulk fermions in the spinor representations of $SO(10)$ are introduced in five dimension, with the SM matter fields identified as zero modes. 
By separating the left- and right-handed chiral fields into different spinor representations, proton decay is structurally forbidden, as baryon-number violating operators cannot be formed. Three Higgs multiplets, including $H_{\bf120}$, $H_{\bf16}$, and $H_{\bf10}$, are introduced to split the masses between quark and lepton masses. In particular, the spinor Higgs $H_{\bf16}$ accounts for the smallness of the light neutrino mass via inverse seesaw.

Based on the framework outlined above, we compute the one-loop RGEs for both gauge and Yukawa couplings. In the flavour sector, we only consider the Yukawa RGEs for the heaviest family ($t$, $b$, $\tau$) in the main text for simplicity, while the full three-family treatment is presented in the appendix. Above the KK threshold, both gauge and Yukawa couplings are formulated in terms of the corresponding 't Hooft couplings, which encode the cumulative contribution from the large number of KK degrees of freedom active below the running scale. The resulting $\beta$ function, written in terms of 't Hooft couplings, include contributions from both zero mode and the KK excitations.

In the gauge sector, the RG running is governed by a linear term proportional to $2\pi$, and a quadratic term involving the $\beta$ coefficient $b_{\bf10}$, defined in Eq.~\eqref{eq:b_10}, for the $SO(10)$ gauge coupling. 
Asymptotic gauge unification is naturally achieved when $b_{\bf10}$, since all gauge 't Hooft couplings approach a universal UV fixed point at $-2\pi/b_{\bf10}$ as the energy scale runs toward the ultraviolet limit, independent of their initial values. For the particle content adopted in this work, this fixed point is given by $6\pi/19$. Note that the requirement for a negative $b_{\bf10}$ restricts the inclusion of redundant Higgs representations in $SO(10)$ GUTs. In particular, the $\overline{\bf126}$-plet Higgs, which is commonly used to fit data for fermion masses and mixing, introduces a large number of degrees of freedom and is thus disfavored in this framework, as it tends to drive $b_{\bf10}$ positive. Nevertheless, viable configurations remain. The Higgs content chosen in this work represents a minimal and phenomenologically viable setup that successfully accounts for fermion mass hierarchies.

Yukawa couplings exhibit a qualitatively different RG running behaviour compared to gauge couplings. Their $\beta$ functions receive contributions from both gauge couplings and Yukawa couplings, with the Yukawa terms always contributing positively. 
We identify a phase diagram of Yukawa couplings featuring a distinct region of asymptotically free phase, a region dominated by a Landau pole in the UV, and two isolated solutions on the separatrix representing asymptotically safe behaviour. The latter are disregarded due to their high degree of  fine-tuning.
Achieving asymptotic freedom requires that the negative gauge contributions outweigh the positive Yukawa self-interactions.  The inclusion of two Higgs fields, $H_{\bf 10}$ and  $H_{\bf 120}$, provides sufficient parameter space to realize this condition. 
Furthermore, due to this different UV behaviour, Yukawa couplings do not exhibit asymptotic unification in the UV. Instead, we require that Yukawa couplings, decomposed from the same Yukawa coupling in SO(10), be explicitly unified at the KK scale. Finally, we demonstrate with a benchmark point that, above the KK scale, the RG flow features a competition between gauge and Yukawa contributions, with the gauge contribution ultimately taking over and driving the 't Hooft Yukawa couplings toward asymptotic freedom.

In conclusion, based on an economical and phenpomenological viable particle content setup, we demonstrate that asymptotic grand unification can be achieved in $SO(10)$ with one extra dimension. While the gauge couplings exhibit asymptotic safety, the Yukawa couplings achieve asymptotic freedom. This work goes beyond the initial study in \cite{Khojali:2022gcq} by discovering that a UV Landau pole phase for the Yukawa couplings can be avoided through the introduction of an additional $H_{\bf 120}$ Higgs multiplet. This work adds a new member to the asymptotic grand unification family, complementing the previous established cases in $SU(5)$ \cite{Cacciapaglia:2020qky} and supersymmetric exceptional $E_6$ case \cite{Cacciapaglia:2023ghp}.

\acknowledgments

ZWW and YLZ give special thanks to Z.Z. Xing for hospitality in IHEP where the initial collaboration was facilitated. They also express their appreciation to G. Cacciapaglia for useful comments on this project, and for providing related computational notes.
GXF and YLZ are partially supported by National Natural Science Foundation of China (NSFC) under Grant Nos. 12205064, 12347103, and Zhejiang Provincial Natural Science Foundation of China under Grant No. LDQ24A050002. ZWW is supported in part by the National Natural Science Foundation of China (Grant No.~12475105).

\appendix

\section{Particle contents and representation decomposition}\label{app:1}

In this appendix, we discuss the decomposition of the particle representations listed in Table~\ref{tab:particle_contents} under the breaking of the $SO(10)$ gauge symmetry to the Pati-Salam subgroup $G_{422}$ and subsequently to the SM gauge group $G_{\rm SM}$.  All decomposed fields are listed in Table~\ref{tab:particle_contents_2}. In both $G_{422}$ and $G_{\rm SM}$, fields denoted by lowercases refer to those have zero modes, while those denoted by uppercases contain only massive KK modes. The BCs for the zero-mode fields (i.e.~with lowercases) on the UV and IR branes are assigned as follows:
\begin{eqnarray}
		\text{left-handed fermions, scalars} &\Rightarrow& (+,+) \,, \nonumber\\
		\text{right-handed fermions} &\Rightarrow& (-,-) \,. 
\end{eqnarray}
The BCs for the KK-only (uppercase) fields are arranged accordingly following the structures of the Yukawa couplings.

Although ${\bf10}$ and ${\bf120}$ are both real representations of $SO(10)$, we arranged $H_{\bf10}$ as a complex scalar and ${\bf120}$ as real, based on a comprehensive consideration of minimality and phenomenological viability. The different arrangement leads to dstinct properties for the resulting fields at lower scale after $SO(10)$ is broken to its subgroups. 
The complex $H_{\bf10}$ decomposes into a bi-doublet Higgs $h_1 \sim (1,2,2)$ under $G_{422}$, and an adjoint Higgs $H_6 \sim (6,1,1)$ under $SU(4)_c$. Since $H_{\bf10}$ is complex, both components are also complex: $h_1 \neq \epsilon h_1^* \epsilon$, where $\epsilon = {\rm i}\sigma_2$; and $H_6 \neq H_6^*$.
Under the SM gauge group $G_{\rm SM}$, the bi-doublet $h_1$ is further decomposed into two complex Higgs doublets $h_u$ and $h_d$, transforming as $h_u, h_d \sim (1, 2, \pm 1/2)$ with $h_u \neq \epsilon h_d^*$. Similarly, $H_6$ yields color triplets satisfying $H_{(3,1,-1/3)} \neq H_{(\bar3,1,1/3)}^*$. 
In contrast, taking $H_{\bf120}$ to be real leads to $h_1'$ and $h_{15}$ being self-conjugate bi-doublets under $SU(2)_L \times SU(2)_R$, satisfying $h_1^{\prime*} =  \epsilon h_1'  \epsilon$ and $h_{15}^* =  \epsilon h_{15}  \epsilon$ respectively. After decomposition under $G_{\rm SM}$, they yield doublets related by complex conjugation, such as $h_u' = \epsilon h_d^{\prime *}$ and $h_u'' = \epsilon h_d^{\prime\prime *}$. Moreover, a real $H_{\bf120}$ implies that both $H_{6L}$ and $H_{6R}$ are real scalar fields, and that $H_{\overline{10}}$ is the complex conjugate of $H_{10}$. 

\begin{table}[h]
	\begin{center}
		\begin{tabular}{ c c c c}
			\hline \hline
			Fields & $SO(10)$ & $G_{422}$ & $G_{\rm SM}$  \\
			\hline
			Fermion & $\Psi_{\bf 16}$ & $\psi_L=(q_L, l_L)_{(4, 2, 1)}$ & 
			${q_L}_{(3,2, \frac16)} + {l_L}_{(1, 2, -\frac12)}$  \\
			& & $\Psi_R^c =(Q_R^c, L_R^c)_{(\overline{4}, 1, 2)}$ & ${D_R^c}_{(\overline{3},1, \frac13)}+{U_R^c}_{(\overline{3},1, -\frac23)}+{N_R^c}_{(1,1, 0)}+{E_R^c}_{(1,1, 1)} $ \\\cline{2-4}
            & $\Psi_{\overline{\bf 16}}$ & $\Psi_L^c =(Q_L^c, L_L^c)_{(\overline{4}, 2, 1)}$ & 
            ${Q_L^c}_{({\bar3}, 2, -\frac16)} + {L_L^c}_{(1, 2, \frac12)}$  \\
            & & $\psi_R=(q_R, l_R)_{(4, 1, 2)}$ & ${d_R}_{(3, 1, -\frac13)}+{u_R}_{(3, 1, \frac23)}+{\nu_R}_{(1, 1, 0)}+{e_R}_{(1, 1, -1)} $ \\\cline{2-4}

			& $\nu_{\rm S}$ & $\nu_{{\rm S}(1, 1, 1)}$ & $\nu_{{\rm S}(1, 1, 0)}$  \\
			\hline
			Higgs & $H_{\bf 10}$ & $h_{1}=(h_u,h_d)_{(1, 2, 2)} $ & ${h_u}_{(1, 2, \frac12)} +{h_d}_{(1, 2, -\frac12)}$ \\
			 &  & $H_{6} =H_{(6, 1, 1)} $ & $H_{(3, 1, -\frac13)} +H_{({\bar3}, 1, \frac13)}$ \\
			\cline{2-4}
			& $H_{\bf 120}$ & $h'_{1}=(h'_u,h'_d)_{(1, 2, 2)}$ & ${h'_u}_{(1, 2, \frac12)} +{h'_d}_{(1, 2, -\frac12)}$ \\
			&  & $h_{15}=(h''_u,h''_d)_{(15,2,2)}$ & ${h''_u}_{(1, 2, \frac12)} +{h''_d}_{(1, 2, -\frac12)}$ \\
			&  & $H_{6L} =H_{(6,3,1)}$ & $H_{(3,3,{-\frac13})}+H_{({\bar3},3,{\frac13})}$ \\
			&  & $H_{6R} =H_{(6,1,3)}$ & $H_{(3, 1, \frac23)} + H_{(3, 1, -\frac13)} + H_{(3, 1, -\frac43)}$ \\ 
			&  &  & $H_{({\bar3}, 1, -\frac23)}+H_{({\bar3}, 1, \frac13)} + H_{({\bar3}, 1, \frac43)}$ \\
			&  & $H_{10} = H_{(10,1,1)}$ & $H_{(1, 1, -1)} + H_{(3, 1, -\frac13)} + H_{(6, 1, \frac13)}$ \\ 
			&  & $H_{\overline{10}} = H_{({\overline{10}},1,1)}$ & $H_{(1, 1, 1)} + H_{({\bar3}, 1, \frac13)} + H_{({\bar6}, 1, -\frac13)}$  
\\\cline{2-4}
			& $H_{\bf 16}$ & 
			$h_{\bar4}=h_{({\bar4}, 1, 2)}$ & $h_{{\rm S}(1,1, 0)} + h_{(1,1, 1)}+h_{({\bar 3},1, -\frac23)} + h_{({\bar3},1, \frac13)}$ \\
			& &  $H_{4}=H_{(4, 2, 1)}$ & 
			$H_{(3,2, \frac16)} + H_{(1,2, -\frac12)}$  \\\hline \hline
		\end{tabular}
	\end{center}
	\caption{Particle contents of $SO(10)$ and their decomposition in the Pati-Salam gauge symmetry $G_{422}$ and SM gauge symmetry $G_{\rm SM}$. The capitalized letters such as $Q,\,L,\,H$ represent fields without zero modes. \label{tab:particle_contents_2}}
\end{table}

\section{Yukawa RGEs with three families in 4D}\label{app:2}

In the Standard Model, each quark and lepton has three families, leading to flavor-dependent Yukawa couplings. Once the family structure is taken into account, the Yukawa couplings for $y_t$, $y_b$ and $y_\tau$  appear as $3\times 3$ matrices. We represent these matrices as $Y_u$, $Y_d$, and $Y_l$, corresponding to the up-type quarks, down-type quarks, and charged leptons, respectively. We further include the $3\times 3$ coefficient matrix $\kappa$ appearing in the dimension-five Weinberg operator:
\begin{align}
	\kappa_{\alpha\beta}[\overline{l_{\alpha L}}\tilde{H}\tilde{H}^T l_{\beta L}^c] + {\rm h.c.}
	\label{eq:Weinberg_operator}
\end{align}
to account for neutrino mass generation in the RG running.
The one-loop RGEs for the Yukawa matrices $Y_u$, $Y_d$, $Y_l$ and $\kappa$ from the EW scale to the PS scale are given by
\begin{align}
	16\pi^2 \frac{{\rm d}Y^{}_{u}}{{\rm d}t}\bigg|_{\rm SM} =& \left[\eta^{}_{u} + \frac{3}{2} (Y^{}_{u} Y^\dagger_{u}) - \frac{3}{2} (Y^{}_{d} Y^\dagger_{d})\right] Y^{}_{u} \,, \nonumber\\
	16\pi^2 \frac{{d}Y^{}_{d}}{{\rm d}t}\bigg|_{\rm SM} =& \left[ \eta^{}_{d} - \frac{3}{2} (Y^{}_{u} Y^\dagger_{u}) + \frac{3}{2} (Y^{}_{d} Y^\dagger_{d})\right]
	Y^{}_{d} \,, \nonumber\\
	16\pi^2 \frac{{\rm d}Y^{}_l}{{\rm d}t}\bigg|_{\rm SM} =& \left[ \eta^{}_l + \frac{3}{2} (Y^{}_lY^\dagger_l) \right] Y^{}_l \,, \nonumber\\
	16\pi^2 \frac{{\rm d}\kappa}{{\rm d}t}\bigg|_{\rm SM} =&  \eta^{}_\kappa \kappa - \frac{3}{2} \left[(Y^{}_l Y^\dagger_l) \kappa + \kappa (Y^{}_lY^\dagger_l)^T  \right] \,. 
	\label{eq:Yukawa_RGE_SM3}
\end{align}
Here, $\eta_u$, $\eta_d$, $\eta_l$, and $\eta_\kappa$ represent flavor-independent correction to the overall magnitude of the coupling matrices, incorporating both gauge suppression and Yukawa self-coupling effects. In the framework of SM we have
\begin{align}
	&\eta^{}_{u} = - \frac{17}{20} g^2_1 - \frac{9}{4} g^2_2 - 8 g^2_3 +
	{\rm Tr} \left[ 3 (Y^{}_{u} Y^\dagger_{u}) +
	3 (Y^{}_{d} Y^\dagger_{d}) + (Y^{}_l Y^\dagger_l) \right] \,, \nonumber \\
	&\eta^{}_{d} = -\frac{1}{4} g^2_1 - \frac{9}{4} g^2_2 - 8 g^2_3 +
	{\rm Tr} \left[ 3 (Y^{}_{u} Y^\dagger_{u}) +
	3 (Y^{}_{d} Y^\dagger_{d}) + (Y^{}_l Y^\dagger_l) \right] \,, \nonumber \\
	&\eta^{}_l = -\frac{9}{4} g^2_1 -\frac{9}{4} g^2_2 +
	{\rm Tr} \left[ 3 (Y^{}_{u} Y^\dagger_{u}) +
	3 (Y^{}_{d} Y^\dagger_{d}) + (Y^{}_l Y^\dagger_l) \right] \,, \nonumber \\
	&\eta^{}_\kappa = -3 g^2_2 + 4\lambda + 
	2{\rm Tr} \left[ 3 (Y^{}_{u} Y^\dagger_{u}) +
	3 (Y^{}_{d} Y^\dagger_{d}) + (Y^{}_l Y^\dagger_l) \right] \,, 
	\label{eq:Yukawa_RGE_coefficientsSM}
\end{align}
where $\lambda$ is the Higgs self-coupling parameter of the SM. 

For energy scale above $M_{\rm PS}$ and below the compactification scale $M_{\rm KK}$, the one-loop RGEs for Yukawa matrices $Y_1$, $Y_1'$, $Y_{15}$, and $Y_4$ can be derived from the Yukawa Lagrangian in Eq.~\eqref{eq:Lagrangian_PS} as follows:
\begin{align}
	16\pi^2\frac{{\rm d}Y_1}{{\rm d}t}=&\eta_1 Y_1+2Y_{1}Y_{1}^{\dagger}Y_{1}+2Y'_{1}Y_{1}^{\dagger}Y'_{1}-\frac{15}{2}Y_{15}Y_{1}^{\dagger}Y_{15}+Y_{1}Y_{1}^{\prime\dagger}Y_{1}^{\prime}+Y_{1}^{\prime}Y_{1}^{\prime\dagger}Y_{1}\nonumber\\
	&+\frac{15}{4}(Y_{1}Y_{15}^{\dagger}Y_{15}+Y_{15}Y_{15}^{\dagger}Y_{1}) 
	+ \frac{1}{2}Y_{1}Y_{4}^{\dagger}Y_{4} \,, \nonumber\\
	16\pi^2\frac{{\rm d}Y_1^\prime}{{\rm d}t}=&\eta'_1Y_{1}^{\prime}+4Y'_1Y_1^{\prime\dagger}Y'_1-\frac{15}{2}Y_{15}Y_{1}^{\prime\dagger}Y_{15}+Y_{1}^{\prime}Y_{1}^{\dagger}Y_{1}+Y_{1}Y_{1}^{\dagger}Y_{1}^{\prime}\nonumber\\
	&+\frac{15}{4}(Y_{1}^{\prime}Y_{15}^{\dagger}Y_{15}+Y_{15}Y_{15}^{\dagger}Y_{1}^{\prime})+\frac{1}{2}Y_{1}^{\prime} Y_{4}^{\dagger} Y_{4} \,, \nonumber\\
	16\pi^2\frac{{\rm d}Y_{15}}{{\rm d}t}=&\eta_{15}Y_{15}-2Y'_{1}Y_{15}^{\dagger}Y'_{1}+7Y_{15}Y_{15}^{\dagger}Y_{15}+Y_{15}Y_{1}^{\dagger}Y_{1}+Y_{1}Y_{1}^{\dagger}Y_{15}+Y_{15}Y_{1}^{\prime\dagger}Y_{1}^{\prime}+Y_{1}^{\prime}Y_{1}^{\prime\dagger}Y_{15}\nonumber\\
	&+\frac{1}{2} Y_{1}Y_{4}^{\dagger}Y_{4}  \,, \nonumber\\
	16\pi^2\frac{{\rm d}Y_{4}}{{\rm d}t}=&\eta_4Y_4+\frac{9}{2}Y_{4}Y_{4}^{\dagger}Y_{4}+(Y_{4}Y_{1}^{\dagger}Y_{1}+Y_{4}Y_{1}^{\prime\dagger}Y_{1}^{\prime})+\frac{15}{4}Y_{4}Y_{15}^{\dagger}Y_{15} \,, 
	\label{eq:Yukawa_RGE_PS3}
\end{align}
where the flavour-independent parts are
\begin{align}
		\eta_1=&-\frac{45}{4}g_{4}^{2}-\frac{9}{4}(g_{2L}^{2}+g_{2R}^{2})+ 4{\rm Tr}(Y_{1} Y_{1}^{\dagger}), \nonumber\\
		\eta'_1=&-\frac{45}{4}g_{4}^{2}-\frac{9}{4}(g_{2L}^{2}+g_{2R}^{2})+ 4{\rm Tr}(Y_{1}^{\prime} Y_{1}^{\prime\dagger}), \nonumber\\
		\eta_{15}=&-\frac{45}{4}g_{4}^{2}-\frac{9}{4}(g_{2L}^{2}+g_{2R}^{2})+{\rm Tr}(Y_{15} Y_{15}^{\dagger}), \nonumber\\
		\eta_4=&-\frac{45}{8}g_{4}^{2}-\frac{9}{4}g_{2L}^{2}+{\rm Tr}(Y_{4} Y_{4}^{\dagger})\,.
\end{align}
Since $h_1$ and $h_1'$ originate from different Higgs multiplets, $H_{\bf 10}$ and $H_{\bf 120}$ respectively, their corresponding Yukawa coupling, $Y_1$ and $Y_1^{\prime}$, arise from distinct representations in $SO(10)$ decomposition. As a result, their products do not form self-coupling terms in the trace, and hence ${\rm Tr}(Y_1^{\prime} Y_1^{\prime\dag})$ does not contribute to $\eta_1$. This is because Yukawa self-coupling contributions typically arise from interactions within the same multiplet, where the quantum numbers match exactly, which is not the case here.
In this analysis, we include the two bi-doublet Higgs fields $h_{1}$ and $h_{1'}$, one colour-triplet Higgs fields $h_{\overline{4}}$, as well as an ajoint Higgs $h_{15}$ in the RG running. If any of these scalars, or their superpositions, have masses above the running energy scale, their contributions in the running should be omitted accordingly. This assumption is consistently applied throughout this work.
For simplicity, we focus on the third-generation Yukawa couplings in the main text. Under this assumption, each Yukawa coupling in Eq.~\eqref{eq:Yukawa_RGE_PS3} is treated as a single number, leading to the following one-loop RGEs:
\begin{align}
		2\pi\frac{{\rm d}\alpha_{y1}}{{\rm d}t}=&\Big[6\alpha_{y1} + 4\alpha_{y1'} + \frac{1}{2}\alpha_{y4} -\frac{45}{4} \alpha_4 -\frac{9}{4}( \alpha_{2L} + \alpha_{2R} )\Big] \alpha_{y1} \,, \nonumber\\
		2\pi\frac{{\rm d}\alpha_{y1'}}{{\rm d}t}=&\Big[ 2\alpha_{y1} + 8\alpha_{y1'} + \frac{1}{2}\alpha_{y4} -\frac{45}{4} \alpha_4 -\frac{9}{4}( \alpha_{2L} + \alpha_{2R} )\Big] \alpha_{y1'} \,, \nonumber\\
		2\pi\frac{{\rm d}\alpha_{y15}}{{\rm d}t}=&\Big[ 8\alpha_{y15} + 2\alpha_{y1} + \frac{1}{2}\alpha_{y4} - \frac{45}{4} \alpha_4 -\frac{9}{4}( \alpha_{2L} + \alpha_{2R} ) \Big] \alpha_{y15} \,, \nonumber\\
		2\pi\frac{{\rm d}\alpha_{y4}}{{\rm d}t}=&\Big[\alpha_{y1} + \alpha_{y1'} +\frac{15}{4} \alpha_{y15} +\frac{11}{2} \alpha_{y4} -\frac{45}{8} \alpha_4 -\frac{9}{4} \alpha_{2L} \Big] \alpha_{y4} \,.
\end{align}
The RGEs for $\alpha_{y1}$, $\alpha_{y1'}$ and $\alpha_{y15}$ in Eq.~\eqref{eq:Yukawa_RGE_PS} are obtained by neglecting the $\alpha_{y4}$ terms in the above equation, which is a reasonable approximation if $\alpha_{y4}\ll 16\alpha_{y1}, 16 \alpha_{y1'}$ is satisfied. 

\section{Calculation of one-loop Yukawa RGEs in 5D SO(10)} \label{app:3}
$SO(10)$ group is defined as the set of linear transformations acting on a 10-dimensional coordinate space $x_a$ for $a=1,2,...10$, which requires the quadratic form $x_1^2+x_2^2+...+x_{10}^2$ to be invariant. It can be written as
\begin{eqnarray}
	x_1^2+x_2^2+...+x_{10}^2=(\Gamma_1x_1+\Gamma_2x_2+...+\Gamma_{10}x_{10})^2\,.
\end{eqnarray}
To satisfy this invariance, the  $\Gamma_a$ matrices (for $a=1,2,...,10$) must anticommute with each other and form a Clifford algebra, defined by
\begin{eqnarray}
	\{\Gamma_a,\Gamma_b\}=2\delta_{ab}\,.
\end{eqnarray}

We follow the constructions of $\Gamma$ matrices in \cite{Rajpoot:1980xy}, which is explicitly written as \cite{DiLuzio:2011mda}, 
\begin{align}
		&\Gamma_1 = \sigma_1\times\sigma_1\times I_2 \times I_2 \times\sigma_2 \,,\hspace{5mm} 
		\Gamma_2 = \sigma_1\times\sigma_2\times I_2\times \sigma_3 \times\sigma_2 \,, \nonumber\\
		&\Gamma_3 = \sigma_1\times\sigma_1\times I_2 \times \sigma_2 \times\sigma_3 \,,\hspace{4.5mm} 
		\Gamma_4 = \sigma_1\times\sigma_2\times I_2 \times \sigma_2 \times I_2 \,, \nonumber\\
		&\Gamma_5 = \sigma_1\times\sigma_1\times I_2\times \sigma_2 \times\sigma_1 \,,\hspace{4.5mm} 
		\Gamma_6 = \sigma_1\times\sigma_2\times I_2 \times \sigma_1 \times\sigma_2 \,, \nonumber\\
		&\Gamma_7 = \sigma_1\times\sigma_3\times \sigma_1\times I_2 \times I_2 \,,\hspace{5mm} 
		\Gamma_8 = \sigma_1\times\sigma_3\times \sigma_2\times I_2 \times I_2 \,, \nonumber\\
		&\Gamma_9 = \sigma_1\times\sigma_3\times \sigma_3\times I_2 \times I_2 \,,\hspace{5mm} 
		\Gamma_{10} = \sigma_2\times I_2 \times I_2 \times I_2 \times I_2 \,,
\end{align}
where the matrices $\sigma_i$ (for $i=1,2,3$) are the standard Pauli matrices and the $\times$ symbol represents the tensor product. 
It is convenient to verify that all the $\Gamma$ matrices constructed above are block off-diagonal, and can be expressed in the form
\begin{align}
	\Gamma_{a} = \begin{pmatrix} 0 & \square \\ \square & 0 \end{pmatrix} \,.
	\label{eq:gamma_matrix}
\end{align}
When the $SO(10)$ group is broken to $SO(6) \times SO(4)$ by boundary conditions,
this basis has the advantage that the first six coordinates $x_1,x_2, ..., x_6$ belong to $SO(6)\simeq SU(4)_c$, while the last four coordinates $x_7,x_8,x_9,x_{10}$ correspond to $SO(4)\simeq SU(2)_L \times SU(2)_R$.  
This 32-dimensional spinor representation is reducible via the chiral projector matrix $\Gamma_\chi$ defined as
\begin{eqnarray}
	\Gamma_\chi=i\Gamma_1\Gamma_2...\Gamma_{10} = \begin{pmatrix} -I_{16} & 0 \\ 0 & I_{16} \end{pmatrix}\,.
\end{eqnarray}
Note that $\Gamma_\chi$ anticommutes with all $\Gamma_a$ matrices i.e.~$\{\Gamma_\chi,\Gamma_a\}=0$ for $a=1,...,10$. This property allows the $32$-dimensional spinor $\Psi$ to be decomposed into two chiral components, which are assigned to the $\Psi_{\bf16}$ and $\Psi_{\overline{\bf16}}$ representations: 
\begin{eqnarray}
	\Psi_{\bf L} \equiv P_{\bf L} \Psi = \begin{pmatrix} \Psi_{\bf 16} \\ 0 \end{pmatrix},\quad 
	\Psi_{\bf R} \equiv P_{\bf R} \Psi = \begin{pmatrix} 0 \\ \Psi_{\overline{\bf 16}} \end{pmatrix}\,,
	\label{eq:32to16}
\end{eqnarray}
where the chiral projection matrices are defined as: $P_{\bf L} = \frac{1}{2} (I_{32} - \Gamma_\chi)$ and $P_{\bf R} = \frac{1}{2} (I_{32} + \Gamma_\chi)$. 
Therefore, the $32$-dimensional spinor representation is reduced into two 16-dimensional irreducible representations, resulting in the decomposition ${\bf 32} = {\bf 16} + \overline{\bf 16}$. The Gauge indices of vector representation (${\bf 45}$) are connected to the spinor indices of the fermion representation (${\bf 32}$) through 45 generators of the $SO(10)$ Lie algebra, defined as
\begin{eqnarray}
	S_{ab}=\frac{i}{4}[\Gamma_a,\Gamma_b] \,,
\end{eqnarray}
where the indices $a,b$ satisfy the antisymmetric relation. Under a small $SO(10)$ rotation, the fermion field $\Psi_i$ transforms as $\Psi_i \to \Psi_i - \frac{i}{2} \lambda_{ab} (S_{ab})_{ij}\Psi_j$. With the help of chiral projection matrices, the generators $S_{ab}$ can be further decomposed into two $16\times 16$ matrices: 
\begin{align}
S_{ab} = P_{\bf L} S_{ab} P_{\bf L} + P_{\bf R} S_{ab} P_{\bf R} =
\frac12 \begin{pmatrix} 
\sigma_{ab} & 0 \\ 0 & 0
\end{pmatrix}
+ \frac12 \begin{pmatrix} 
0 & 0 \\ 0 & \tilde{\sigma}_{ab} 
\end{pmatrix}\,,
\end{align}
where $\sigma_{ab}$ and $\tilde{\sigma}_{ab}$ are the representation matrices of ${\bf 45}$ in the spinor representations ${\bf 16}$ and $\overline{\bf 16}$, respectively. Under a small  $SO(10)$ rotation, $\Psi_{\bf L}$ and $\Psi_{\bf R}$ transform as 
$(\Psi_{\bf L})_i \to (\Psi_{\bf L})_i - \frac{i}{4} \lambda_{ab} (\sigma_{ab})_{ij}(\Psi_{\bf L})_j$ and 
 $(\Psi_{\bf R})_i \to (\Psi_{\bf R})_i - \frac{i}{4} \lambda_{ab} (\tilde{\sigma}_{ab})_{ij}(\Psi_{\bf R})_j$.

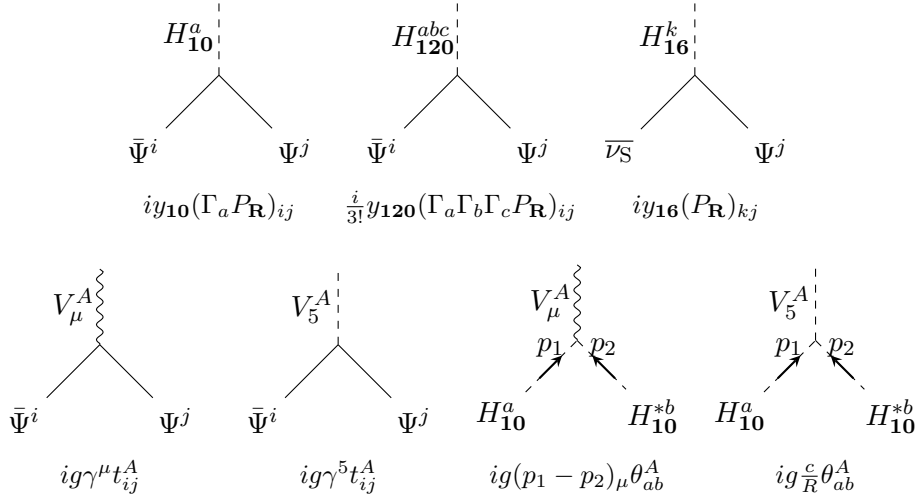
\begin{figure}[htbp]
	\centering
	\begin{subfigure}[b]{0.2\textwidth}
		\centering
		\begin{tikzpicture}
			\begin{feynman}
				\vertex (a) at (0, -1) {$\bar{\Psi}^i$};
				\vertex (b) at (1, 1) ;
				\vertex (c) at (2, -1) {$\Psi^j$};
				\vertex (d) at (1, 0);
				\vertex (e) at (1, 0.15);
				\vertex (f) at (1,0.65);
				\vertex (h) at (0.6,0.5) {$H_{\bf 10}^{a}$};
				\diagram* {
					(a) -- [plain] (d) -- [plain] (c),
					(d) -- [scalar] (b),
				};
			\end{feynman}
		\end{tikzpicture}
		\caption*{$i y_{\bf10}(\Gamma_{a}P_{\bf R})_{ij}$}
	\end{subfigure}
	\begin{subfigure}[b]{0.2\textwidth}
		\centering
		\begin{tikzpicture}
			\begin{feynman}
				\vertex (a) at (0, -1) {$\bar{\Psi}^i$};
				\vertex (b) at (1, 1) ;
				\vertex (c) at (2, -1) {$\Psi^j$};
				\vertex (d) at (1, 0);
				\vertex (e) at (1, 0.15);
				\vertex (f) at (1,0.65);
				\vertex (h) at (0.55,0.5) {$H_{\bf 120}^{abc}$};
				\diagram* {
					(a) -- [plain] (d) -- [plain] (c),
					(d) -- [scalar] (b),
				};
			\end{feynman}
		\end{tikzpicture}
		\caption*{$\frac{i}{3!}y_{\bf120}(\Gamma_a\Gamma_b\Gamma_c P_{\bf R})_{ij}$}
	\end{subfigure}
	\begin{subfigure}[b]{0.2\textwidth}
		\centering
		\begin{tikzpicture}
			\begin{feynman}
				\vertex (a) at (0, -1) {$\overline{\nu_{\rm S}}$};
				\vertex (b) at (1, 1) ;
				\vertex (c) at (2, -1) {$\Psi^j$};
				\vertex (d) at (1, 0);
				\vertex (e) at (1, 0.15);
				\vertex (f) at (1,0.65);
				\vertex (h) at (0.6,0.5) {$H_{\bf 16}^{k}$};
				\diagram* {
					(a) -- [plain] (d) -- [plain] (c),
					(d) -- [scalar] (b),
				};
			\end{feynman}
		\end{tikzpicture}
		\caption*{$iy_{\bf16}(P_{\bf R})_{kj}$}
	\end{subfigure} \\[5mm]
	\begin{subfigure}[b]{0.2\textwidth}
		\centering
		\begin{tikzpicture}
			\begin{feynman}
				\vertex (a) at (0, -1) {$\bar{\Psi}^i$};
				\vertex (b) at (1, 1) ;
				\vertex (c) at (2, -1) {$\Psi^j$};
				\vertex (d) at (1, 0);
				\vertex (e) at (1, 0.15);
				\vertex (f) at (1,0.65);
				\vertex (h) at (0.65,0.5) {$V_\mu^A$};
				\diagram* {
					(a) -- [plain] (d) -- [plain] (c),
					(d) -- [photon] (b),
				};
			\end{feynman}
		\end{tikzpicture}
		\caption*{$ig\gamma^\mu t^{A}_{ij}$}
	\end{subfigure}
	\begin{subfigure}[b]{0.2\textwidth}
		\centering
		\begin{tikzpicture}
			\begin{feynman}
				\vertex (a) at (0, -1) {$\bar{\Psi}^i$};
				\vertex (b) at (1, 1) ;
				\vertex (c) at (2, -1) {$\Psi^j$};
				\vertex (d) at (1, 0);
				\vertex (e) at (1, 0.15);
				\vertex (f) at (1,0.65);
				\vertex (h) at (0.65,0.5) {$V_5^A$};
				\diagram* {
					(a) -- [plain] (d) -- [plain] (c),
					(d) -- [scalar] (b),
				};
			\end{feynman}
		\end{tikzpicture}
		\caption*{$ig\gamma^5 t^{A}_{ij}$}
	\end{subfigure}
	\begin{subfigure}[b]{0.2\textwidth}
		\centering
		\begin{tikzpicture}
			\begin{feynman}
				\vertex (a) at (0, -1) {$H_{\bf 10}^a$};
				\vertex (b) at (1, 1) ;
				\vertex (c) at (2, -1) {$H_{\bf 10}^{* b}$};
				\vertex (d) at (1, 0);
				\vertex (e) at (1, 0.15);
				\vertex (f) at (1,0.65);
				\vertex (h) at (0.65,0.5) {$V_\mu^A$};
				\diagram* {
					(a) -- [scalar] (d) -- [scalar] (c),
					(d) -- [photon] (b),
				};
			\end{feynman}
			\draw[->, thick, >=stealth] (0.5,-0.5) -- (0.8,-0.2) node[midway, above] {$p_1$};
			\draw[->, thick, >=stealth] (1.5,-0.5) -- (1.2,-0.2) node[midway, above] {$p_2$};
		\end{tikzpicture}
		\caption*{$ig(p_1-p_2)_\mu \theta^{A}_{ab}$}
	\end{subfigure}
	\begin{subfigure}[b]{0.2\textwidth}
		\centering
		\begin{tikzpicture}
			\begin{feynman}
				\vertex (a) at (0, -1) {$H_{\bf 10}^a$};
				\vertex (b) at (1, 1) ;
				\vertex (c) at (2, -1) {$H_{\bf 10}^{* b}$};
				\vertex (d) at (1, 0);
				\vertex (e) at (1, 0.15);
				\vertex (f) at (1,0.65);
				\vertex (h) at (0.65,0.5) {$V_5^A$};
				\diagram* {
					(a) -- [scalar] (d) -- [scalar] (c),
					(d) -- [scalar] (b),
				};
			\end{feynman}
			\draw[->, thick, >=stealth] (0.5,-0.5) -- (0.8,-0.2) node[midway, above] {$p_1$};
			\draw[->, thick, >=stealth] (1.5,-0.5) -- (1.2,-0.2) node[midway, above] {$p_2$};
		\end{tikzpicture}
		\caption*{$ig\frac{c}{R} \theta^{A}_{ab}$}
	\end{subfigure}
	\caption{Feynman rules of Yukawa vertex and gauge boson vertex.}
	\label{fig:Feynman_rules}
\end{figure}

In our model, the $SO(10)$-invariant Yukawa terms in Eq.~\eqref{eq:Lagrangian_SO10} are written out explicitly as
\begin{align}
	-\mathcal{L}_Y=& \overline{P_{\bf L}\Psi}\,( y_{\bf10} \Gamma_{a} H_{\bf 10}^a+ \frac{1}{3!} y_{\bf120}\Gamma_a\Gamma_b\Gamma_c H_{\bf 120}^{abc}) \,P_{\bf R}\Psi 
	+y_{\bf16}\overline{\nu_{\rm S}}H_{\bf 16} \, P_{\bf R} \Psi + {\rm h.c.}
	\nonumber\\
	=& \bar{\Psi}^i \, \big(y_{\bf10}(\Gamma_{a}P_{\bf R})_{ij}H_{\bf 10}^a+ \frac{1}{3!}y_{\bf120}(\Gamma_a\Gamma_b\Gamma_c P_{\bf R})_{ij}H_{\bf 120}^{abc} \big) \, \Psi^j 
	+y_{\bf16}\overline{\nu_{\rm S}}H_{\bf 16}^k (P_{\bf R})_{kj} \Psi^j + {\rm h.c.}\,, 
	\label{eq:SO(10)Yukawa}
\end{align}
where the indices run as $i,j = 1,2,...,32$ for spinors and $a,b,c=1,...,10$ with $a \neq b \neq c$, and the Higgs field $H_{\bf16}$ is embedded in the ${\bf 32}$ spinor representation as $(H_{\bf 16}, 0)^T$. Indices $a, b,c$ in ${\bf120}$ Higgs are fully antisymmetric, which requires a normalization factor of $(3!)^{-1}$ to account for the number of possible permutations. The product of $\Gamma$ matrices establishes the connection between the gauge indices of the Higgs fields and the spinor indices of the fermions, e.g., $\overline{P_{\bf L} \Psi} \,  \Gamma_a \, P_{\bf R} \,\Psi = \bar{\Psi}^i \,  (\Gamma_a \, P_{\bf R})_{ij} \,\Psi^j$. 
Based on these terms, we can derive Feynman rules of Yukawa vertex and gauge boson vertex directly, as shown in Fig.~\ref{fig:Feynman_rules}. 

To compute all possible diagrams contributing to one-loop Yukawa RGEs, we utilize the following key identities for the $\Gamma$ matrices: $\{\Gamma_a,\Gamma_b\}=2\delta_{ab},\ \Gamma_a\Gamma_a=I_{32}$, ${\rm Tr}(\Gamma_a\Gamma_a)=32$ and ${\rm Tr}(\Gamma_a\Gamma_a P_{\bf R})=16$ ($a$ is not summed). Extending the analysis from a single generation to three generations in flavour space, the Yukawa couplings $y_{\bf10}$, $y_{\bf120}$ and $y_{\bf16}$ become $3\times 3$ Yukawa matrices, denoted as $Y_{\bf10}$, $Y_{\bf120}$, and $Y_{\bf16}$. Since all fermions are arranged in 16-dimensional spinor representation, we can directly derive the one-loop Yukawa RGEs in usual 4-dimensional spacetime by eliminating the contribution of $V_5$, which is the fifth component of the gauge field. Additionally, the factor $c$ appearing in the final diagram is an effective KK-mode-dependednt coefficient obtained after integrating over profiles in extra dimension, inversely proportional to the KK mass. As a result, the contribution of these diagrams is suppressed for large KK modes and can be safely neglected in the one-loop RG running. When calculating group theoretic factors, the computation involves the anticommutation relations of the $\Gamma$ matrices.  Specifically, the Feynman rules for Yukawa vertices involving the ${\bf 120}$-plet Higgs require the multiplication of three $\Gamma$ matrices. Some representative Feynman diagrams for the one-loop Yukawa RGEs of $Y_{\bf120}$ are displayed in Table.~\ref{tab:diagrams}. 
\begin{table}[t!] 
	\renewcommand{\arraystretch}{1.8}
	\centering
	\begin{tabular}{ |m{3cm}<{\centering} | m{7cm}<{\centering} |}
		\hline
		Diagrams & Group theoretic factor \\
		\hline
		\begin{tikzpicture}
			\begin{feynman}
				\vertex (a) at (0, -1){$\bar{\Psi}^i$};
				\vertex (b) at (1, 1) {$H_{\bf 120}^{abc}$};
				\vertex (c) at (2, -1) {$\Psi^j$};
				\vertex (d) at (1, 0);
				\vertex (e) at (0.6, -0.4);
				\vertex (f) at (1.4,-0.4);
				\vertex (g) at (0.25,-0.15) {$\Psi^k$};
				\vertex (h) at (1.75,-0.15) {$\bar{\Psi}^l$};
				\vertex (k) at (1,-0.8) {$H_{\bf 120}^{def}$};
				\diagram* {
					(a) -- [plain] (d) -- [plain] (c),
					(d) -- [scalar] (b),
					(e) -- [scalar] (f),
				};
			\end{feynman}
		\end{tikzpicture} & $(\Gamma_{def}P_{\bf R})_{lj}(\Gamma_{cba}P_{\bf R})_{kl}(\Gamma_{def}P_{\bf R})_{ik} =(\Gamma_{def}\Gamma_{cba}\Gamma_{def}P_{\bf R})_{ij} = -8(\Gamma_{cba} P_{\bf R})_{ij} =8(\Gamma_{abc} P_{\bf R})_{ij}$ 
		\\ \hline 
		\begin{tikzpicture}
		\begin{feynman}
			\vertex (a) at (0, -1){$\bar{\Psi}^i$};
			\vertex (b) at (1, 1) {$H_{\bf 120}^{abc}$};
			\vertex (c) at (2, -1) {$\Psi^j$};
			\vertex (d) at (1, 0);
			\vertex (e) at (1, 0.1);
			\vertex (f) at (1,0.6);
			\vertex (g) at (0.3,-0.1) {$H_{\bf 120}^{def}$};
			\vertex (h) at (0.3,0.4) {$\bar{\Psi}^k$};
			\vertex (k) at (1.7, 0.4) {$\Psi^l$};
			\diagram* {
				(a) -- [plain] (d) -- [plain] (c),
				(b) -- [scalar] (f),
				(e) -- [plain,half left] (f) -- [plain,half left] (e),
				(e) -- [scalar] (d),
			};
		\end{feynman}
		\end{tikzpicture} & $(\Gamma_{def} P_{\bf R})_{ij}(\Gamma_{fed} P_{\bf R})_{kl}(\Gamma_{abc} P_{\bf R})_{lk}=\frac{1}{36}3!(\Gamma_{def} P_{\bf R})_{ij} \times{\rm Tr}(\Gamma_f\Gamma_e\Gamma_d\Gamma_a\Gamma_b\Gamma_c P_{\bf R})=\frac{1}{36}(3!)^2(\Gamma_{abc} P_{\bf R})_{ij} \times 16=16(\Gamma_{abc} P_{\bf R})_{ij}$ \\
		\hline
		\begin{tikzpicture}
			\begin{feynman}
				\vertex (a) at (0, -1){$\bar{\Psi}^i$};
				\vertex (b) at (1, 1) {$H_{\bf 120}^{abc}$};
				\vertex (c) at (2, -1) {$\Psi^j$};
				\vertex (d) at (1, 0);
				\vertex (e) at (0.85, -0.15);
				\vertex (f) at (0.5,-0.5);
				\vertex (g) at (0.,-0.3) {$\Psi^k$};
				\vertex (h) at (0.5,0.1) {$\bar{\Psi}^l$};
				\vertex (k) at (1.1,-0.7) {$H_{\bf 120}^{def}$};
				\diagram* {
					(a) -- [plain] (d) -- [plain] (c),
					(d) -- [scalar] (b),
					(e) -- [scalar,half left] (f),
				};
			\end{feynman}
		\end{tikzpicture} & $(\Gamma_{abc} P_{\bf R})_{lj}(\Gamma_{fed} P_{\bf R})_{kl}(\Gamma_{def} P_{\bf R})_{ik}= \frac{1}{36}3!(\Gamma_f\Gamma_e\Gamma_d\Gamma_d\Gamma_e\Gamma_f\Gamma_{abc} P_{\bf R})_{ij}=\frac{10\times9\times8}{6}(\Gamma_{abc} P_{\bf R})_{ij}=120(\Gamma_{abc} P_{\bf R})_{ij}$ 
		\\ \hline 
		\begin{tikzpicture}
		\begin{feynman}
			\vertex (a) at (0, -1){$\bar{\Psi}^i$};
			\vertex (b) at (1, 1) {$H_{\bf 120}^{abc}$};
			\vertex (c) at (2, -1) {$\Psi^j$};
			\vertex (d) at (1, 0);
			\vertex (e) at (1.5, -0.5);
			\vertex (f) at (1.15,-0.15);
			\vertex (g) at (1.9,-0.3) {$\bar{\Psi}^l$};
			\vertex (h) at (1.5,0.1) {$\Psi^k$};
			\vertex (k) at (0.85,-0.8) {$H_{\bf 120}^{def}$};
			\diagram* {
				(a) -- [plain] (d) -- [plain] (c),
				(d) -- [scalar] (b),
				(e) -- [scalar,half left] (f),
			};
		\end{feynman}
		\end{tikzpicture} & $(\Gamma_{def} P_{\bf R})_{lj}(\Gamma_{fed} P_{\bf R})_{kl}(\Gamma_{abc} P_{\bf R})_{ik}= \frac{1}{36}3!(\Gamma_{abc}\Gamma_d\Gamma_e\Gamma_f\Gamma_f\Gamma_e\Gamma_d P_{\bf R})_{ij}=\frac{10\times9\times8}{6}(\Gamma_{abc} P_{\bf R})_{ij}=120(\Gamma_{abc} P_{\bf R})_{ij}$ \\
		\hline
	\end{tabular}
	\caption{Part of Feynman diagrams related to one-loop Yukawa RGEs of $Y_{\bf120}$. Corresponding group theoretic factor is also displayed. Here $\Gamma_{abc}$ is a short notation of $\frac{1}{3!} \Gamma_a \Gamma_b \Gamma_c$ with $a \neq b \neq c$. Indices appearing in internal propagators are summed. }
	\label{tab:diagrams}
\end{table}

One-loop RGEs for any $3\times 3$ Yukawa coupling matrix $Y_{\bf r}$ above $M_{\rm KK}$ take the form
\begin{equation}
		16\pi^2\frac{{\rm d}Y_{\rm r}}{{\rm d}t} = 16\pi^2\frac{{\rm d}Y_{\rm r}}{{\rm d}t}\bigg|_{4\rm D}+(S(t)-1)16\pi^2\frac{{\rm d}Y_{\rm r}}{{\rm d}t}\bigg|_{\rm KK} \,,
	\label{eq:YukawaED}
\end{equation}
where $S(t)$ is the same as in Eq.~\eqref{eq:S(t)}. Adding all contributions of the possible diagrams, explicit formulas of $\frac{{\rm d}Y_{\rm r}}{{\rm d}t}\big|_{\rm 4D}$ and $\frac{{\rm d}Y_{\rm r}}{{\rm d}t}\big|_{\rm KK}$ for Yukawa matrices appearing in Eq.~\eqref{eq:SO(10)Yukawa} are given below
\begin{align}
		16\pi^2\frac{{\rm d}Y_{\bf 10}}{{\rm d}t}\bigg|_{\rm 4D}=&\eta_{\bf10}\big|_{\rm 4D}Y_{\bf10}-96Y_{\bf120}Y_{\bf10}^{\dagger}Y_{\bf120}+10Y_{\bf10}Y_{\bf10}^{\dagger}Y_{\bf10}\nonumber\\
		&+60(Y_{\bf10}Y_{\bf120}^{\dagger}Y_{\bf120}+Y_{\bf120}Y_{\bf120}^{\dagger}Y_{\bf10})+\frac{1}{2}Y_{\bf10}Y_{\bf16}^{\dagger}Y_{\bf16} \,, \nonumber\\
		16\pi^2\frac{{\rm d}Y_{\bf120}}{{\rm d}t}\bigg|_{\rm 4D}=&\eta_{\bf120}\big|_{\rm 4D}Y_{\bf120}+104Y_{\bf120}Y_{\bf120}^{\dagger}Y_{\bf120}+5(Y_{\bf120}Y_{\bf10}^{\dagger}Y_{\bf10}+Y_{\bf10}Y_{\bf10}^{\dagger}Y_{\bf120}) \nonumber\\
		&+\frac{1}{2}Y_{\bf120}Y_{\bf16}^{\dagger}Y_{\bf16} \,, \nonumber\\
		16\pi^2\frac{{\rm d}Y_{\bf16}}{{\rm d}t}\bigg|_{\rm 4D}=&\eta_{\bf16}\big|_{\rm 4D}Y_{\bf16}+\frac{17}{2}Y_{\bf16}Y_{\bf16}^{\dagger}Y_{\bf16}+5Y_{\bf16}Y_{\bf10}^{\dagger}Y_{\bf10}+60Y_{\bf16}Y_{\bf120}^{\dagger}Y_{\bf120} \,, 
		\label{eq:Yukawa_RGE_SO10_4D}
\end{align}
	and
\begin{align}
		16\pi^2\frac{{\rm d}Y_{\bf10}}{{\rm d}t}\bigg|_{\rm KK}=&\eta_{\bf10}\big|_{\rm KK}Y_{\bf10}-96Y_{\bf120}Y_{\bf10}^{\dagger}Y_{\bf120}+10Y_{\bf10}Y_{\bf10}^{\dagger}Y_{\bf10}\nonumber\\
		&+60(Y_{\bf10}Y_{\bf120}^{\dagger}Y_{\bf120}+Y_{\bf120}Y_{\bf120}^{\dagger}Y_{\bf10})+\frac{1}{2}Y_{\bf10}Y_{\bf16}^{\dagger}Y_{\bf16} \,, \nonumber\\
		16\pi^2\frac{{\rm d}Y_{\bf120}}{{\rm d}t}\bigg|_{\rm KK}=&\eta_{\bf120}\big|_{\rm KK}Y_{\bf120}+104Y_{\bf120}Y_{\bf120}^{\dagger}Y_{\bf120}+5(Y_{\bf120}Y_{\bf10}^{\dagger}Y_{\bf10}+Y_{\bf10}Y_{\bf10}^{\dagger}Y_{\bf120}) \nonumber\\
		&+\frac{1}{2}Y_{\bf120}Y_{\bf16}^{\dagger}Y_{\bf16} \,, \nonumber\\
		16\pi^2\frac{{\rm d}Y_{\bf16}}{{\rm d}t}\bigg|_{\rm KK}=&\eta_{\bf16}\big|_{\rm KK}Y_{\bf16}+\frac{17}{2}Y_{\bf16}Y_{\bf16}^{\dagger}Y_{\bf16}+5Y_{\bf16}Y_{\bf10}^{\dagger}Y_{\bf10}+60Y_{\bf16}Y_{\bf120}^{\dagger}Y_{\bf120} \,, 		\label{eq:Yukawa_RGE_SO10_KK}
\end{align}
where
\begin{align}
		&\eta_{\bf10}\big|_{\rm 4D}=-\frac{270}{8}g_{\bf10}^2+8\; {\rm Tr}(Y_{\bf10}Y_{\bf10}^{\dagger}) \,, \nonumber\\
		&\eta_{\bf120}\big|_{\rm 4D}=-\frac{270}{8}g_{\bf10}^2+8\; {\rm Tr}(Y_{\bf120}Y_{\bf120}^{\dagger}) \,, \nonumber\\
		&\eta_{\bf16}\big|_{\rm 4D}=-\frac{135}{8}g_{\bf10}^2+{\rm Tr}(Y_{\bf16}Y_{\bf16}^{\dagger}) \,, 
\end{align}
and
\begin{align}
		&\eta_{\bf10}\big|_{\rm KK}=-\frac{171}{8}g_{\bf10}^2+16\;{\rm Tr}(Y_{\bf10}Y_{\bf10}^{\dagger}) \,, \nonumber\\
		&\eta_{\bf120}\big|_{\rm KK}=-\frac{219}{8}g_{\bf10}^2+16\; {\rm Tr}(Y_{\bf120}Y_{\bf120}^{\dagger}) \,, \nonumber\\
		&\eta_{\bf16}\big|_{\rm KK}=-\frac{225}{16}g_{\bf10}^2+2\; {\rm Tr}(Y_{\bf16}Y_{\bf16}^{\dagger}) \,.
\end{align}
Note that above the KK sclae, all Yukawa contributions in trace terms are doubled compared to their 4D counterparts. This enhancement arises from the simultaneous presence of both $\Psi_L$ and $\Psi_R$ components in the fermion loop.

\section{RGEs matching between PS and SO(10) above KK scale}\label{app:4}
An SO(10)-invariant theory can be reformulated within the PS framework. Based on the field decomposition provided in Table.~\ref{tab:particle_contents_2}, the explicit form of the SO(10)-invariant Yukawa couplings is given by 
\begin{align}
y_{\bf 10} \overline{\Psi_{\bf 16}} H_{\bf 10} \Psi_{\overline{\bf 16}} =& 		
 y_1 \big( \overline{\psi_L} h_1 \psi_R + \overline{\Psi_R^c} h_1 \Psi_L^c \big) + y_6 \big( \overline{\psi_L} H_6 \Psi_L^c +  \overline{\Psi_R^c} H_6 \psi_R \big) \,.
\nonumber\\
{\rm{i}}\, y_{\bf 120} \overline{\Psi_{\bf 16}} H_{\bf 120} \Psi_{\overline{\bf 16}} 
=& y_1' \big( \overline{\psi_L} h_1' \psi_R + \overline{\Psi_R^c} h_1' \Psi_L^c \big)
+ y_{15} \big( \overline{\psi_L} h_{15} \psi_R + \overline{\Psi_R^c} h_{15} \Psi_L^c \big) \nonumber\\
&+ y_6' ( \overline{\psi_L} H_{6L} \Psi_L^c +  \overline{\Psi_R^c} H_{6R} \psi_R ) + y_{10} ( \overline{\psi_L} H_{10} \Psi_L^c +  \overline{\Psi_R^c} H_{\overline{10}} \psi_R ) \,, \nonumber\\
y_{\bf 16} \overline{\nu_{\rm S}} H_{\bf 16} \Psi_{\overline{\bf 16}} =& y_4 \overline{\nu_{\rm S}} h_{\bar4} \psi_R + y_{4}' \overline{\nu_{\rm S}} H_4 \Psi_L^c \,.
\end{align}
The matching relations between Yukawa couplings in $SO(10)$ and PS symmetries, as derived in \cite{Aulakh:2002zr}, are 
\begin{align}
 &y_{\bf10} = \frac{1}{\sqrt{2}} y_1 = \frac{1}{2} y_6 \,, \quad 
 y_{\bf16} = y_4=y_4' \,, \nonumber\\
 &y_{\bf120} = \frac{1}{\sqrt{2}}y_1' = \frac{1}{2\sqrt{2}} y_{15} = \frac{1}{4} y_6' = \frac{1}{2\sqrt{2}} y_{10} \,. \label{eq:matching_SO10_PS}
\end{align}
The RGEs of $y_1$, $y_1'$, $y_{15}$, and $y_{4}$ above the KK scale are derived below: 
\begin{align}
		16\pi^2\frac{{\rm d}y_1}{{\rm d}t}\Big|_{\rm KK} =&\Big[10 y_1^2 +\frac32 y_6^2 + 4 y_1^{\prime 2} -\frac52 y_{10}^2 +\frac94 y_6^{\prime 2} + \frac{1}{2} y_4^2 \nonumber\\
		&  - \frac{81}{8} g_4^2 - \frac{45}{8} (g_{2L}^2 + g_{2R}^2 )\Big] y_1 \,, \nonumber\\
		16\pi^2\frac{{\rm d}y_1'}{{\rm d}t}\Big|_{\rm KK} =&\Big[2 y_1^2 + \frac{3}{2} y_6^2 + 12 y_1^{\prime 2} +\frac{15}{2} y_{10}^2 + \frac{9}{4}y_6^{\prime 2} + \frac{1}{2} y_4^2  \nonumber\\
		& - \frac{129}{8} g_4^2 - \frac{45}{8}( g_{2L}^2 + g_{2R}^2 )\Big] y_1' \,, \nonumber\\
		16\pi^2\frac{{\rm d}y_{15}}{{\rm d}t}\Big|_{\rm KK} =&\Big[ 2 y_1^2 + \frac32 y_6^2 + 9 y_{15}^2 + \frac32 y_{10}^2 + \frac94 y_6^{\prime 2} + \frac{1}{2} y_4^2  \nonumber\\
		& - \frac{129}{8} g_4^2 - \frac{45}{8}( g_{2L}^2 + g_{2R}^2 ) \Big] y_{15} \,, \nonumber\\
		16\pi^2\frac{{\rm d}y_4}{{\rm d}t}\Big|_{\rm KK} =&\Big[ y_1^2 + \frac34 y_6^2 + y_1^{\prime 2} + \frac{15}{4} y_{15}^2 + \frac54 y_{10}^2 + \frac98 y_6^{\prime 2} + \frac{21}{2} y_4^2 \nonumber\\
		&- \frac{135}{16} g_4^2 - \frac{45}{8} g_{2R}^2 \Big] y_4 \,.
	\label{eq:Yukawa_RGE_KKp}
\end{align}
When the matching condition with $y_{\bf10}$, $y_{\bf120}$, and $y_{\bf4}$ in Eq.~\eqref{eq:matching_SO10_PS} are satisfied, the above RGEs naturally reduce to the $SO(10)$ Yukawa RGE given in Eq.~\eqref{eq:Yukawa_RGE_SO10_KK}. It is noteworthy that, while the Yukawa RGEs for $y_1'$ and $y_1$ take the same form in 4D, they exhibit slight differences above the KK scale. This discrepancy arises due to the additional interactions between $h_1'$ and the $(6,2,2)$ gauge boson present only above the KK scale.

\end{document}